\documentclass[review]{elsarticle}
\usepackage[utf8]{inputenc}
\usepackage{url}

\usepackage{lineno}

\usepackage[a4paper, total={6in, 8in}]{geometry}

\usepackage{etoolbox}
\makeatletter
\usepackage{comment}

\patchcmd{\ps@pprintTitle}
  {\let\@oddhead\@empty}
  {\def\@oddhead{\mbox{}\hfill\thepage}}
  {}{}
\makeatother

\usepackage{psfrag,epsfig,graphics}
\usepackage{amsmath,amsthm,amssymb,multirow}
\usepackage{mathbbol}
\usepackage{amssymb}             

\DeclareSymbolFontAlphabet{\amsmathbb}{AMSb}%
\usepackage{mathabx} 

\usepackage{booktabs}
\usepackage{graphicx,amsmath,graphicx}
\usepackage{caption}

\usepackage{multirow}
\usepackage[lined,linesnumbered,ruled]{algorithm2e}

\usepackage{color}  

\newcommand{\cb}[1]{{\boldsymbol{#1}}}
\newcommand{\cp}[1]{\ifmmode {\mathcal{#1}}\else ${\mathcal{#1}}$\fi}

\newcommand{\bB}{\boldsymbol{B}}
\newcommand{\bC}{\boldsymbol{C}}

\newcommand{\bH}{\boldsymbol{H}}
\newcommand{\bI}{\boldsymbol{I}}
\newcommand{\bK}{\boldsymbol{K}}
\newcommand{\bL}{\boldsymbol{L}}

\newcommand{\bP}{\boldsymbol{P}}

\newcommand{\bR}{\boldsymbol{R}}
\newcommand{\bS}{\boldsymbol{S}}

\newcommand{\bX}{\boldsymbol{X}}

\newcommand{\bh}{\boldsymbol{h}}

\newcommand{\be}{\boldsymbol{e}}

\newcommand{\bq}{\boldsymbol{q}}
\newcommand{\bo}{\boldsymbol{o}}
\newcommand{\br}{\boldsymbol{r}}
\newcommand{\by}{\boldsymbol{y}}
\newcommand{\bs}{\boldsymbol{s}}
\newcommand{\bu}{\boldsymbol{u}}

\newcommand{\bx}{\boldsymbol{x}}

\newcommand{\bz}{\boldsymbol{z}}

\newcommand{\calD}{\mathcal{D}}

\newcommand{\calH}{\mathcal{H}}
\newcommand{\calG}{\mathcal{G}}
\newcommand{\calL}{\mathcal{L}}
\newcommand{\calN}{\mathcal{N}}

\usepackage[euler]{textgreek}
\usepackage[colorinlistoftodos]{todonotes}

\newcommand{\bpi}{\boldsymbol{\pi}}
\newcommand{\bmu}{\boldsymbol{\mu}}
\newcommand{\btheta}{\boldsymbol{\theta}}

\newcommand{\bsigma}{\boldsymbol{\sigma}}

\newcommand{\bLambda}{\boldsymbol{\Lambda}}

\newcommand{\bSigma}{\boldsymbol{\Sigma}}

\newcommand{\bxi}{\boldsymbol{\xi}}

\newcommand{\Ex}{\amsmathbb{E}}
\newcommand{\vect}{\operatorname{vec}}
\newcommand{\diag}{\operatorname{diag}}
\newcommand{\blkdiag}{\operatorname{blkdiag}}

\usepackage{color}  

\definecolor{darkgreen}{rgb}{0., 0.4, 0.}
\definecolor{amber}{rgb}{1.0, 0.49, 0.0}
\definecolor{orange}{rgb}{1.0, 0.4, 0.0}
\definecolor{darkorange}{rgb}{0.7, 0.32, 0.0}

\begin{document}

\makeatletter
\def\ps@pprintTitle{%
   \let\@oddhead\@empty
   \let\@evenhead\@empty
   \let\@oddfoot\@empty
   \let\@evenfoot\@empty
}
\makeatother

\begin{frontmatter}

\title{Robust Recursive Fusion of Multiresolution Multispectral Images \\ with Location-Aware Neural Networks}

\author[1]{Haoqing Li} 

\affiliation[1]{organization={University of Calgary},
            addressline={2500 University Drive N.W}, 
            city={Calgary},
            postcode={T3A2V9}, 
            state={AB},
            country={Canada}}

\author[2]{Ricardo Borsoi} 

\affiliation[2]{organization={University of Lorraine, CNRS},
            addressline={CRAN}, 
            city={Nancy},
            postcode={F-54000}, 
            state={},
            country={France}}

\author[3]{Tales Imbiriba}

\affiliation[3]{organization={University of Massachusetts Boston},
            addressline={100 Morrissey Blvd}, 
            city={Boston},
            postcode={02125}, 
            state={MA},
            country={USA}}
\author[4]{Pau Closas}

\affiliation[4]{organization={Northeastern University},
            addressline={360 Huntington Avenue}, 
            city={Boston},
            postcode={02115}, 
            state={MA},
            country={USA}}

\fntext[fn1]{E-mail: 
haoqing.li@ucalgary.ca, ricardo-augusto.borsoi@cnrs.fr, 
tales.imbiriba@umb.edu, pau.closas@northeastern.edu}

\begin{abstract}
Multiresolution image fusion has been studied for years to solve the trade-off between temporal and spatial resolution in remote sensing instruments, and has been widely applied to detect and monitor natural phenomena like floods. Despite the considerable research on this topic, the consideration of mitigating outliers influence in satellite image fusion, such as cloud and shadow miscorrections, is not fully developed. Moreover, strategies that integrate robustness, recursive operation and learned models are missing. In this paper, we designed a robust recursive image fusion framework leveraging location-aware neural networks (NN) to model the image dynamics is proposed. Outliers are modeled by representing the probability of contamination of a given pixel and band. A NN model trained on a small dataset provides accurate predictions of the stochastic image time evolution, which improves both the accuracy and robustness of the method. A recursive solution is proposed to estimate the high-resolution images using a Bayesian variational inference framework. Experiments fusing images from the Landsat~8 and MODIS instruments show that the proposed approach is significantly more robust against cloud cover, without losing performance when no clouds are present.
\end{abstract}




\end{frontmatter}

\section{Introduction}

Satellite-based remote sensing is an essential tool for many applications such as the monitoring of land-cover use~\cite{lu2016land}, deforestation~\cite{schultz2016performance} or water
quality~\cite{gholizadeh2016comprehensive}.
A major concern when leveraging satellite-based imaging regards the trade-off among temporal, spatial, and spectral resolutions. 
Such trade-off is due to the large distances from space-borne instruments and target scenes, and limitations of multiband imaging systems. These limitations imply that higher spatial resolution leads to longer revisit times.
For instance, the Moderate Resolution Imaging Spectroradiometer (MODIS) has pixel sizes of 250/500/1000 $m$ with daily revisit period while Landsat~8 captures images with pixel sizes of 30/100 $m$ once every 16 days~\cite{roy2014landsat}. 
This is a major issue when monitoring events that change rapidly and at the same time require high spatial resolution to be properly characterized.

To cope with these limitations image fusion approaches were proposed to generate high-spatial-temporal resolution images by fusing image data from multiple space-borne instruments~\cite{wang2018sentinel2_sentinel3_fusion} to generate daily high (e.g., 30 $m$) resolution images~\cite{yang2021forestMortality_HSR,rittger2021MEF_snowcover_30m}. 
Spatial-temporal image fusion approaches can be divided into four categories. Unmixing-based algorithms decompose the images into the spectral signatures of the constituent materials and their pixelwise abundances, which reduces the dimensionality of the problem~\cite{li2020sfsdaf_enhanced_spatiotemporal_fusion,Borsoi_2018_Fusion,huang2012spatiotemporalFusionDictLearning,borsoi2023learningDisentangledUnmixing}.
Weighted fusion predicts the temporal changes in the high spatial resolution images using a weighted linear combination of the temporal changes of low spatial resolution pixels in the observed area~\cite{gao2006fusion_MODIS_denseTimeSeries_exp,zhu2010fusion_ESTARFM,zhang2018spatiotemporalFusionWeightedMultiscale}.
Bayesian approaches can model the uncertainty of satellite images and further estimate high-resolution images
~\cite{xu2021landTemperatureFusionKalman,li2023online}. Finally, learning-based approaches using neural networks (NN) leverage some kind of training data to learn a mapping from the low-resolution to the high resolution images~\cite{song2018spatiotemporalFusionCNNs,tan2021flexibleFusionImagesGANs,benzenati2024stf_transformerImageFusion,wang2023deepUnmixingInterImageVariability}. 

Despite the efficiency of these methods, outliers caused by effects such as cloud contamination can severely impact their image fusion performance. Thus, the detection and removal of pixels contaminated by clouds is an essential step of image fusion pipelines. Different algorithms for cloud (and cloud shadow) detection and removal have been proposed, which
can be divided into three categories~\cite{wang2023virtual}. The first kind restores the cloudy/shadow contaminated pixels by assuming they share the same statistical distribution or geometric structures as the surrounding cloudless ones~\cite{siravenha2011evaluating,xu2022attention}. 
The second kind uses auxiliary information from different sensors, such as synthetic aperture radar \cite{meraner2020cloud} or MODIS images \cite{shen2019spatiotemporal}. The third kind uses cloudless images from the same sensor on other dates as reference images \cite{
zhang2020thick}.

However, the quality of cloud removal is limited, and the images used in the fusion process might still contain outliers.
Thus, the development of robust image fusion methods is paramount. 
%
Moreover, although noise-robust image fusion methods have been proposed~\cite{tan2022robustFusionMODIS_Landsat}, there is a lack of robust online image fusion approaches.

Recursive image fusion methods process the low-resolution image sequence on the fly to generate the high-resolution images, benefiting from the information contained in long sequences while retaining a fixed computational complexity. 
The Kalman filter framework~\cite{sarkka2013bayesianBook} has been adopted in several works as it provides a statistically principled solution to this problem. Such methods were developed to estimate time series of vegetation indices~\cite{sedano2014kalmanFUsionNDVI} and land surface temperature~\cite{xu2021landTemperatureFusionKalman} and also to fuse Landsat and MODIS images in an offline manner~\cite{zhou2020kalmanFusionLandsatMODIS}.
A key element of this framework is a model for the time evolution of the high-resolution images. This is related to a multispectral video prediction problem. Recently, a Bayesian filtering approach for image fusion was proposed in~\cite{li2023online} using a linear and Gaussian model for the image evolution whose covariance was estimated from small amounts of historical data. The flexibility of this model, however, is very limited. For computer vision, several works proposed black-box NN-based approaches for video prediction
\cite{oprea2020reviewDeepVideoPrediction,gao2022simvpSimpleCNNVideoPrediction,rakhimov2020latentVideoTransformer,wang2022predrnn_RNN_videoPrediction,voleti2022conditionalVideoDiffusionPredictionGeneration}. 
However, such approaches require large amounts of training data.

\begin{figure}[t]
    \centering
    \includegraphics[width=0.8\linewidth]{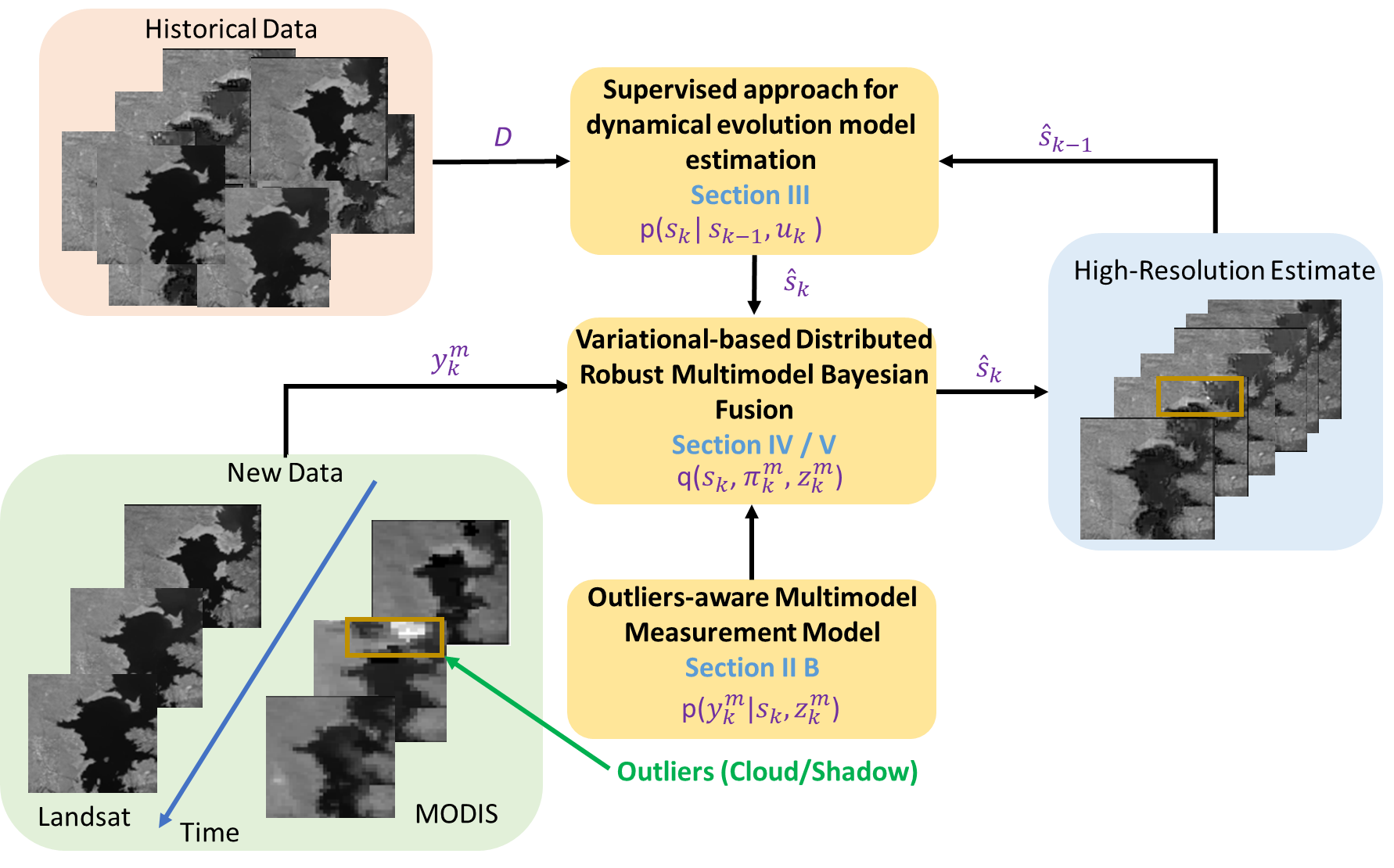}\\
    \vspace{-0.15cm}
    \caption{\footnotesize Overview of the proposed method. Time series of multimodal (e.g., Landsat and MODIS) images are recursively fused by the proposed robust Variational Distributed Multimodal Bayesian Fusion algorithm. The fused results are high spatial-temporal resolution estimated images, while the contamination caused by cloud/shadow influence (marked in yellow box) is mitigated. A location-aware NN-based dynamical evolution model is estimated through a supervised strategy based on a small amount of high-resolution historical data.}
    \label{fig:overview}
\end{figure}

In this paper we propose a robust recursive image fusion framework using location-aware NNs to tackle the aforementioned limitations. An illustration of the proposed method is shown in Figure~\ref{fig:overview}.
First, a probabilistic image acquisition model is presented in Section~\ref{sec:model} where the measurement outliers such as clouds are represented as a statistical hypothesis of a large-magnitude outlier affecting a pixel, not requiring a rigid statistical model.
Then, in Section~\ref{sec:NN} a location aware NN-based stochastic model is proposed to represent the dynamical evolution of the high-resolution images. The NNs, which represent the mean and variance of the image evolution, are based on interpretable structures that use auxiliary variables to aid in the predictions. The NNs are trained on a small dataset of historical images of a given scene within a maximum-likelihood framework. Images from different modalities are then recursively fused in a Bayesian paradigm leveraging a variational inference framework previously developed in~\cite{li2020robust} discussed in Section~\ref{sec:VBKF}. To handle the high computation complexity of the Bayesian solution, a new computationally efficient distributed algorithm is developed in Section~\ref{sec:distributed} by using different statistical approximations. This makes it possible to employ this solution over large datasets and geographical areas. Experimental results with real Landsat-8 and MODIS data shown in Section~\ref{sec:experiments} illustrate the superior performance of the proposed approach.

In summary, the contribution of the proposed method is the design of a Convolutional Neural Network {CNN}-based Bayesian variational inference framework to mitigate the cloud cover influence for a more accurate multiresolution and multispectral image fusion. Furthermore,  the novelty of the proposed method is in a principled estimation framework which combines:
1) robustness to measurement outliers under minimal assumption on their statistical distribution, 2) scalable recursive operation, 3) approximated distributed implementation to reduce computational cost, and 4) integration of location-aware NNs to model image dynamics which require only small amounts of training data.

\section{Dynamical Imaging Model}
\label{sec:model}

\subsection{Definitions and notation}
In this paper, we define $\by_{k,\ell}^{m}\in{\amsmathbb{R}}^{N_{m,\ell}}$ as the $k$-th acquired image reflectances at $\ell$-th band with modality under $m\in\Omega$, $N_{m,\ell}$ pixels for each of the bands $\ell=1,\ldots,L_m$, and $\Omega$ as the set of image modalities. To be specific, we consider Landsat-8 and MODIS images in this paper, which are represented by $\Omega=\{\mathsf{L},\mathsf{M}\}$. 
%
We denote the high-resolution latent reflectances by $\bS_k\in{\amsmathbb{R}}^{N_H\times L_H}$, with $N_H$ pixels and $L_H$ bands, with $L_H\geq L_m$ and $N_H\geq N_{m,\ell}$, $\forall \ell,m$. Subindex $k\in{\amsmathbb{N}}_*$ indicates the acquisition date. Furthermore, the vectorization, vector stacking, diagonal and block diagonal matrix operators are denoted by $\operatorname{vec}(\cdot)$, $\operatorname{col}\{\cdot\}$, $\operatorname{diag}\{\cdot\}$ and by $\operatorname{blkdiag}\{\cdot\}$, respectively. The notation $\bx_{a:b}$ for $a,b\in{\amsmathbb{N}}_*$ represents the set $\{\bx_a,\bx_{a+1},\ldots,\bx_b\}$. ${\calN}(\bmu,\bSigma)$ is the Gaussian distribution, with $\bmu$ as mean, and $\bSigma$ as covariance matrix.

\subsection{Measurement model}
The images acquired at each time $k$ consist of spatially degraded, noisy versions of a high-resolution image $\bS_k$. Following this assumption, traditional methods consider a measurement model that can be expressed according to:
\begin{align}
    \by_{k,\ell}^{m} = {\calH}_{\ell}^{m}\big(\bS_k\big) + \br_{k,\ell}^{m} \,,
    \label{eq:obs_mdl_1}
\end{align}
for each band $\ell=1,\ldots,L_m$ and for each modality $m\in\Omega$. The function ${\calH}_{\ell}^{m}:\amsmathbb{R}^{N_H\times L_H}\to\amsmathbb{R}^{N_{m,\ell}}$ is a linear operator representing the spectral and spatial degradation occurring at the $\ell$-th band of the $m$-th imaging modality; it represents the effects of blurring and downsampling of the high resolution image bands, as well as the spectral response function of the $\ell$-th sensor band. $\br_{k,\ell}^{m}$ is the measurement noise. Note that we assume in this paper different bands in $\bS_k$ holds the same spatial resolution, but the resolution can be different over bands. 
Normally, the measurement noise is assumed as uncorrelated Gaussian noise among band, that is, $\br_{k,\ell}^{m}\sim{\calN}(\cb{0},\bR_{\ell}^m)$, where $\bR_{\ell}^m\in{\amsmathbb{R}}^{N_{m,\ell}\times N_{m,\ell}}$ is the time-invariant covariance matrix for this Gaussian distribution. 
At each time index $k$ the scene is measured through one of the imaging modalities $m\in\Omega$.


Using~\eqref{eq:obs_mdl_1} and properties of the vectorization of matrix products, we can stack all bands of the $m$-th modality in the vector $\by_{k}^{m}\in{\amsmathbb{R}}^{n_y^m}$, with $n_y^m=\sum_{\ell} N_{m,\ell}$ being the total amount of pixels observed in the $m$-th image modality, leading to the equivalent reformulation of model~\eqref{eq:obs_mdl_1} as
\begin{align}
    \by_{k}^{m} = {\bH}^{m}\bs_k + {\br}_{k}^{m} \,,
    \label{eq:meas_mdl_4new}
\end{align}
where $\by_{k}^{m} = \operatorname{col}\big\{{\by}_{k,1}^{m},\ldots,{\by}_{k,L_m}^{m}\big\}$, ${\br}_{k}^{m} = \operatorname{col}\big\{{\br}_{k,1}^{m},\ldots,{\br}_{k,L_m}^{m}\big\}$, ${\bR}^m = \operatorname{blkdiag}\big\{\bR_{1}^m, \ldots, \bR_{L_m}^m\big\}$, ${\bH}^{m}  = \big[ \big(\bH_{1}^{m} \big) ^\top, \cdots, \big(\bH_{L_m}^{m}\big) ^\top\big]^\top$, and $\bH_{\ell}^{m}$ is a matrix form representation of the operator ${\calH}_{\ell}^{m}$, such that $\vect({\calH}_{\ell}^{m}(\bS_k))=\bH_{\ell}^{m}\bs_k$. $\bs_k\in\amsmathbb{R}^{L_H N_H}$ denotes a vector-ordering of the high-resolution image $\bS_k$ which is obtained by reorganizing all pixels to have the bands of a single high-resolution pixel, and positions corresponding to nearby pixels are adjacent to each other (see \cite{li2023online} for details about how the pixel ordering).

In satellite imaging, the pixel values can be degraded by several potential influences in the image fusion process. For example, they can be affected by the dead pixels in the sensor, atmospheric compensation error and the presence of cloud and shadow. Most existing algorithms ignore the existence of such outlier pixels, which can lead to a considerable loss of performance when they are applied in real settings.
%
Thus, we address this issue by considering two hypotheses for the measurements. Under the first hypothesis, denoted by $\mathcal{C}_0$, the pixels are only affected by Gaussian noise ${\br}_{k}^{m}$, whereas under the second hypothesis, denoted by $\mathcal{C}_1$, the pixels are corrupted, being affected by a vector of outliers ${\bo}_{k}^{m}\in\amsmathbb{R}^{n_y^m}$. This leads to the following measurement model:
\begin{align}
   {y}_{k}^{m,(i)} &=\left\{
		\begin{array}{ll}
			{\bh}^{m,(i)}\bs_k + {r}_{k}^{m,(i)}  \,,  &  \text{\,\,\, under } \mathcal{C}_0 \\
			{\bh}^{m,(i)}\bs_k + {r}_{k}^{m,(i)}  + {o}_{k}^{m,(i)} \,,  &  \text{\,\,\, under } \mathcal{C}_1 \\
		\end{array}
   \label{eq:obs_mdl_2}
    \right.
\end{align}
for $i=1,\ldots,n_y^m$, where ${y}_{k}^{m,(i)}$, ${r}_{k}^{m,(i)}$ and ${o}_{k}^{m,(i)}$ denote the $i$-th element of ${\by}_{k}^{m}$, ${\br}_{k}^{m}$ and ${\bo}_{k}^{m}$
respectively, and ${\bh}^{m,(i)}$ denotes the $i$-th row of ${\bH}^{m}$.
Note that we have one hypothesis for each band and pixel in the measurement of modality $m$, which might be affected by an outlier.
Moreover, the approach we will consider will not need a rigid statistical model for ${\bo}_{k}^{m}$, as will be shown in the following section.

\section{Learning a scene-adapted dynamical model}
\label{sec:NN}
To properly exploit the temporal information in the image sequence, an adequate dynamical model is necessary to describe the dynamics of the HR images.
Model-based frameworks aim to construct priors for the dynamical image evolution solely based on prior knowledge, such as, assuming that changes in video sequences are of low magnitude, zero mean and spatially smooth \cite{borsoi2018newVideoSRR}. However, the reconstruction quality achieved using such models is limited.

Recently, the use of data-driven models has become the predominant approach to perform video prediction~\cite{oprea2020reviewDeepVideoPrediction}. 
Such approaches are typically based on end-to-end learnable neural network architectures such as CNNs~\cite{gao2022simvpSimpleCNNVideoPrediction}, RNNs~\cite{wang2022predrnn_RNN_videoPrediction} and transformers~\cite{rakhimov2020latentVideoTransformer,weissenborn2019autoregressiveVideoTransformer}.
The flexibility of such models allows for accounting for long-term temporal dependencies.
Stochastic models for video prediction have also been proposed to account for the uncertainty in the predictions~\cite{babaeizadeh2018stochasticVideoPrediction,castrejon2019improvedVRNNsVideoPrediction,franceschi2020stochasticResidualVideoPrediction}. 
%
These models are trained in a variational inference framework~\cite{castrejon2019improvedVRNNsVideoPrediction,borsoi2023dynamicalVRNNunmixing,Bishop:2006fk}.
More recently, diffusion models have been applied to stochastic video prediction~\cite{voleti2022conditionalVideoDiffusionPredictionGeneration}, owing to their great success in video generation tasks~\cite{blattmann2023videoSynthesisLatentDiffusionModels}. Other learning approaches leveraging physical knowledge through partial differential equations have also been recently explored~\cite{guen2020disentanglingPDEvideoprediction,singh2023learningSemilinear}.



For remote sensing applications, neural network models based on the transformer architecture have become well-established for forecasting tasks~\cite{gao2022earthformer,jiao2023transformerVideoRemoteSensing}.
However, despite their flexibility such black-box models require large amounts of training data and are not easily interpretable.
%
In \cite{li2023online}, a simple data-driven model was proposed based on a linear and Gaussian model with a data-driven selection of innovation covariance matrix, which needs only a small number of historical images.

Existing prediction models for videos are either black-box models which use large amounts of data for training~\cite{wang2022predrnn_RNN_videoPrediction,rakhimov2020latentVideoTransformer}, or rely on analytical models~\cite{li2023online} which are interpretable but too simplistic to properly describe the changes seen in real multispectral image sequences.
To overcome this limitation, we propose a prediction model for the high-resolution images based on location-aware neural networks which incorporates an interpretable architecture design. Moreover, the data-driven parts of the model, consisting of CNNs, are trained using only small amounts of data consisting of a historical dataset $\calD$ of high-resolution images of the scene to be processed. 

\subsection{Supervised dynamical evolution model learning}\label{sec:Qest}


In this section we propose to learn a scene-adapted dynamical model $p_{\phi}(\bs_{k}|\bs_{k-1};\bu_{k})$, which is assumed to be a conditionally Gaussian distribution represented as:
\begin{align}
    & p(\bs_{k}|\bs_{k-1};\bu_{k}) 
    = \calN\big(\bmu_{\phi}\big(\bs_{k-1},\bu_{k}), \operatorname{diag}(\bsigma^2_{\phi}(\bs_{k-1},\bu_{k}))\big) \,,
\end{align}
where vector $\bu_{k}$ contains additional variables used in the prediction, and functions $\bmu_{\phi}$ and $\bsigma^2_{\phi}$ compute the mean and the diagonal of the covariance matrix of the distribution.
This corresponds to the following equivalent dynamical model:
\begin{equation}
     \bs_{k} = \mu_{\phi}(\bs_{k-1},\bu_{k}) + \bq_{k-1} \,,
    \label{eq:state_evol_mdl_2}
\end{equation}
with $\bq_{k-1}\sim{\calN}(\cb{0},\operatorname{diag}(\bsigma^2_{\phi}(\bs_{k-1},\bu_{k})))$.


\subsection{Additional input variables $\bu_k$}

Additional variables provide useful information for video prediction in several tasks such as Atari games~\cite{oh2015actionConditionedVideoPredictionATARI} or robot agents motion~\cite{finn2016videoPredictionRobot}. In remote sensing, environmental variables have been used to improve the estimation of vegetation indices from satellite data~\cite{li2024incorporatingEnviroinmentalVariablesNDVI}.
In this work, to help model the image dynamics we consider a vector of auxiliary deterministic variables, denoted by $\bu_k$ and defined as
\begin{align}
    \bu_k = \big[\boldsymbol{pos}^\top,\, \bq_0^\top,\, date_k,\, \Delta_k \big]^\top\,,
\end{align}
where $date_k$ denotes the date of the $k$-th acquisition (number of days since the first day of the corresponding year for the earliest acquisition date), $\Delta_k$ the number of days between the $k-1$-th and the $k$-th acquisitions, $\boldsymbol{pos}$ is a vector containing the spatial position of each pixel, and $\bq_0$ is the variance of the image difference computed from historical data. It is inspired by the weakly supervised covariance estimation method in \cite{li2023online}, aiming to provide a rough estimation for the average transition rate. To compute it, let us define a dataset of historical images of the scene under consideration by $\calD = \{\by_{i}^{L,h}:i=1,\ldots,n_h\}$ with $n_h$ Landsat images at previous time instants. Based on $\calD$, $\bq_0$ is computed as:
\begin{equation}
    \bq_0 = \frac{1}{n_{\calD}} \frac{1}{\Delta_{\rm median}} \sum_{\bs\in\calD} \Big(\bs - \frac{1}{n_h} \sum_{\bs'\in\calD} \bs' \Big)^2 \,,
    \label{eq:Q0}
\end{equation}
where $\Delta_{\rm median}$ denotes the median time between acquisition dates of consecutive images in $\calD$, which is used to normalize the result to a per-day variance estimate, and $n_{\calD}$ is the total number of historical images in historical dataset, and the squaring in \eqref{eq:Q0} is computed elementwise.

\subsection{NN Model structure}

\begin{figure}
    \centering
    \includegraphics[width = 0.65\columnwidth]{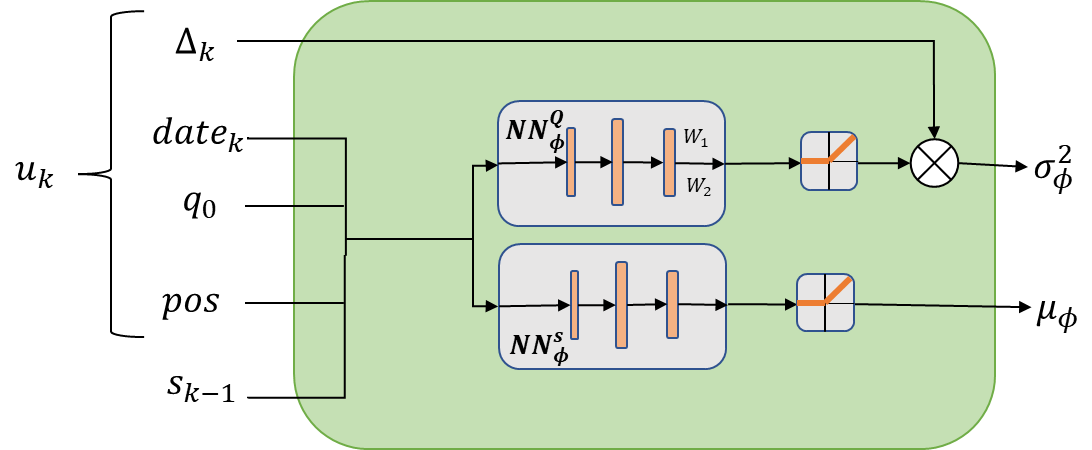}
    \vspace{-0.1cm}
    \caption{Structure of the NN model used to compute the pixelwise mean and variance of the distribution of time evolution of the high-resolution images.}
    \label{fig:DNNSturcture}
\end{figure}

We propose to parameterize the functions $\bmu_{\phi}$ and $\bsigma^2_{\phi}$ by combining an interpretable model with CNNs learned based on the historical dataset. $\bmu_{\phi}$ is given by
\begin{align}
    \bmu_{\phi}\big(\bs_{k-1},\bu_k) = {\rm ReLU}\big(\bs_{k-1} + {\rm NN}_{\phi}^s(\bs_{k-1},\bu_k)\big) \,,
    \label{eq:NN_model_mu}
\end{align}
which is composed by the sum of the image estimate at the previous time instant $\bs_{k-1}$ and a residual computed by the CNN estimate ${\rm NN}_{\phi}^s(\bs_{k-1},\bu_k)$, where $\phi$ represents the parameters of CNN. ${\rm ReLU}(\cdot)$ denotes the rectified linear unit function, which is used to guarantee that the predicted mean is nonnegative. Note that \eqref{eq:NN_model_mu} is a general expression for estimating $\bmu_{\phi}$. The diagonal of the covariance of the predictive distribution $\bsigma^2_{\phi}$ is given by:
\begin{align}
    \bsigma^2_{\phi}(\bs_{k-1},\bu_k) ={} & \Delta_k \times {\rm ReLU} \big(W_1 \times \bq_0
    + W_2 \times {\rm NN}_{\phi}^Q(\bs_{k-1},\bu_k)\big) \,,
    \label{eq:NN_model_sigma}
\end{align}
where $W_1,W_2\in\amsmathbb{R}$ are weighting parameters. 
This is a weighted combination between the variances $\bq_0$ computed a priori based on the historical dataset and a residual term ${\rm NN}_{\phi}^Q(\bs_{k-1},\bu_k)$ computed by a CNN. The ${\rm ReLU}(\cdot)$ function ensures that the computed variances are nonnegative, and the result is scaled by $\Delta_k$ so that the variance becomes larger (resp., smaller) the longer (resp., shorter) the time interval between the acquisitions $k-1$ and $k$.
%
The structure of the NN is shown in Fig.~\ref{fig:DNNSturcture}, where we use ${\rm NN}_{\phi}^s$ and ${\rm NN}_{\phi}^Q$ to represent the neural networks generally. For details about the architecture of the NNs (${\rm NN}_{\phi}^s$ and ${\rm NN}_{\phi}^Q$) and the corresponding input, please refer to Appendix~\ref{app:CNN_architecture}.


\subsection{Cost function}

To learn the parameters of the NN-based models in~\eqref{eq:NN_model_mu} and~\eqref{eq:NN_model_sigma}, we consider two objectives. The main objective is to maximize the expected log-likelihood of the corresponding transition PDF $p(\bs_{k}|\bs_{k-1};\bu_{k})$, given by $\Ex_{p(\bs_{k-1},\bs_{k})} \big\{\log( p(\bs_{k}|\bs_{k-1};\bu_{k}) )\big\}$. When the expectation is approximated using the samples contained in the historical dataset $\calD$, this leads to the following objective:
\begin{align}
     \calL_{\rm LL}(\phi) 
    &={} \frac{1}{n_{\bs}} \sum_{(\bs_{k-1},\bs_{k})\in \calD} \log\, p(\bs_{k}|\bs_{k-1};\bu_{k})
    \nonumber\\
    & ={} \frac{1}{2n_{\bs}} \sum_{(\bs_{k-1},\bs_{k})\in \calD} \bigg[ \big\|\bs_{k}- \bmu_{\phi}\big(\bs_{k-1},\bu_k)\big\|^2_{[\diag(\bsigma^2_{\phi}(\bs_{k-1},\bu_k))]^{-1}} 
    \nonumber \\
    & \qquad\qquad  - \log\,\det\big(\diag(\bsigma^2_{\phi}(\bs_{k-1},\bu_k))\big) + \kappa \bigg] \,,
\end{align}
where $\kappa$ is a constant. Note that, with a slight abuse of notation, we use $\{(\bs_{k-1},\bs_{k})\in \calD\}$ to represent the set of image pairs observed at consecutive time instants in the historical dataset, with $n_{\bs}$ being the number thereof.



Based on $\mathcal{L}_{\rm LL}(\phi)$, we consider the following objective function to train the neural network model:
\begin{equation}
    \calG(\phi,W_1,W_2) = \calL_{\rm LL}(\phi) + \lambda_1\big[ (W_1-1)^2 + W_2^2\big] + \lambda_2 \mathcal{R}(\phi) \,,
    \label{eq:loss_function_prediction_mdl}
\end{equation}
where $\lambda_1,\lambda_2$ are regularization parameters. The first regularization term in \eqref{eq:loss_function_prediction_mdl}, $\big[ (W_1-1)^2 + W_2^2\big]$ penalizes the magnitude of the weight $W_2$ and forces $W_1$ to be close to one, thus, the larger its contribution in the loss term, the more the covariance model in~\eqref{eq:NN_model_sigma} will approximate the a priori estimate given by $\Delta_k\times\bq_0$, reducing the contribution of the  NN. The last term, $\mathcal{R}(\phi)$, is a weight decay which penalizes the L$_1$ norm of the parameters of the neural networks ${\rm NN}_{\phi}^Q$ and ${\rm NN}_{\phi}^s$.


\section{Robust image fusion through variational Kalman filter}
\label{sec:VBKF}
Given the probabilistic model for the image generation and its temporal dynamics, the online image fusion consists of computing the posterior distribution of the high resolution image conditioned on all past measurements, $p\big(\bs_k\big|\{{\by}_{1:k}^{m}\}_{m\in\Omega}\big)$. To simplify the notation, we will suppress the dependence of some distributions on the input variables $\bu_k$ whenever this does not impair clarity.
When the model is linear and Gaussian, this PDF can be computed efficiently using the Kalman filter~\cite{sarkka2013bayesianBook,li2023online}. However, the proposed model tackles the presence of outliers, and uses a learned temporal evolution model, which such techniques cannot address.

To address this issue, we consider an approach based on the general VBKF (GBVKF) proposed in \cite{li2020robust}. This section simply provides an overview of the proposed GBVKF presented in \cite{li2020robust}. First, let us introduce the outlier indicator vector $\bz_k^m = \left(z_{k}^{m,(1)},\dots , z_{k}^{m,(n_y^m)}\right)^\top \in \mathcal{Z}=\{0,1\}^{n_y^m}$, 
such that $z_{k}^{m,(i)}=0$ if there is an outlier on the $i$-th (corrupted) element of ${\by}_{k}^{m}$, i.e., ${y}_{k}^{m,(i)}$, and $z_{k}^{m,(i)}=1$ if the $i$-th element is otherwise clean (not corrupted). The clean {elements} of ${\by}_{k}^{m}$ can be used nominally in the image fusion process, whereas the contribution of the corrupted ones should be down-weighted. {This is performed by modifying the observation model such that the $i$-th position of indicator vector, $z_{k}^{m,(i)}$, adjusts the variance of a modified (referred to as \textit{improper}) Gaussian noise distribution, leading to
\begin{align}
    p\left({\by}_{k}^{m}|\bs_k,\bz_k^m\right) & \,\propto\, \exp\Big({-\frac{1}{2}\big\|{\by}_{k}^{m}- {\bH}^{m}\bs_k\big\|^2_{\bSigma_k^{-1}(\bz_k^m)}}\Big) \,, 
    \label{eq:modified-likelihood}
\end{align}
where each element of the covariance matrix in the $i$-th rwo and $j$-th column is defined as $[\bSigma_k(\bz_k^m)]_{ij} = [{\bR}^m]_{ij}$ for $i \neq j$ and $[\bSigma_k(\bz_k^m)]_{ij} = [{\bR}^m]_{ij}/z_{k}^{m,(i)}$ for $i = j$, where $[{\bR}^m]_{ij}$ is the $ij$-th element of ${\bR}^m$.

To solve the image fusion problem under the new model, we need to approximate the posterior distribution $p\left(\bs_k,\bz_k^m|{\by}_{k}^{m}\right)$. Under the assumption of Bayesian framework, we impose a beta-Bernoulli hierarchical prior to each indicator in $\bz_k^m$, to estimate the posterior of $\bz_k^m$. Specifically, we assume $p\big(z_k^{m,(i)}\big|\pi_k^{m,(i)}\big) = B(z_k^{m,(i)},{\pi_k^{m,(i)}})$ is a Bernoulli distribution and
$\pi_k^{m,(i)}$ follows beta distribution, defined by (unknown) shape hyperparameters
$e_0^{(i)}$ and $f_0^{(i)}$, i.e., $p\big(\pi_k^{m,(i)}\big) = \text{Beta}(e_0^{(i)}, f_0^{(i)})$, for $i=1,\ldots,n_y^m$.

Note that we assume that the indicators are independent to each other in this paper.
To estimate the posterior distribution of the latent variables $\btheta=\{ \bs_k, \bpi_k^m, \bz_k^m\}$, that is  $p(\btheta|{\by}_{1:k}^{m})$, we apply the Variational Inference (VI) principle \cite{Bishop:2006fk, VB_book}, and resort to an auxiliary distribution $q(\btheta)$ using independence assumptions such that:
\begin{equation}
q\left(\btheta\right)=q\left(\bs_k\right)q\left(\bpi_k^m\right)q\left(\bz_k^m\right)=q\left(\bs_k\right)\prod_{i=1}^{n_y^m}q\big(\pi_k^{m,(i)}\big)q\big(z_k^{m,(i)}\big).
\label{q_join_est}
\end{equation}
\noindent  
Then we apply the mean-field VI method to the joint distribution and acquire the various marginal distributions, $q(\cdot)$,
\begin{align} 
    p\big(\bs_k,\bpi_k^m, & \bz_k^m,{\by}_{1:k}^{m}\big) 
    \,\propto\, p\left(\bs_k|{\by}_{1:k-1}^{m}\right) p\left({\by}_{k}^{m}|\bs_k,\bz_k^m\right) p(\bz_k^m , \bpi_k^m), 
\label{eq:jointpdf-gen}
\end{align}
such that $\ln \left[q(\bs_k)\right]$, $\ln \left[q(\bpi_k^m)\right]$ and $\ln \left[q(\bz_k^m)\right]$ can be obtained by marginalizing the logarithm of the joint distribution in~\eqref{eq:jointpdf-gen}.
Within the Gaussian filtering framework, the first term $p\left(\bs_k|{\by}_{1:k-1}^{m}\right)$ on the right-hand side of (\ref{eq:jointpdf-gen}) is a predictive density, which can be approximated as $p\left(\bs_k|{\by}_{1:k-1}^{m}\right) \approx  \mathcal{N}\big(\hat{\bs}_{k|k-1},\widehat{\bP}_{k|k-1}\big)$,
where $\hat{\bs}_{k|k-1}$ and $\widehat{\bP}_{k|k-1}$ are the mean and covariance matrix of $p\left(\bs_k|{\by}_{1:k-1}^{m}\right)$, which can be computed using cubature integration rules~\cite{Aras-09}.

The second and third terms $p\left({\by}_{k}^{m}|\bs_k,\bz_k^m\right) p(\bz_k^m , \bpi_k^m)$ on the right-hand side of (\ref{eq:jointpdf-gen}) correspond to the update phase in a Kalman filter, which can be approximated using \eqref{q_join_est}, with $q(\bs_k)\approx \calN(\hat{\bs}_{k|k},\widehat{\bP}_{k|k})$, where  
$\hat{\bs}_{k|k}$ and $\widehat{\bP}_{k|k}$ are the mean and covariance of the filtering posterior at time $k$, that is $p\left(\bs_{k}|{\by}_{1:k}^{m}\right) \approx \mathcal{N}\big(\hat{\bs}_{k|k},\widehat{\bP}_{k|k}\big)$, while $q(\bz_k^m)$ and $q(\bpi_k^m)$ are approximated as Bernoulli and beta distribution, respectively.
To be detailed, we specify term $p(\bz_k^m , \bpi_k^m)$ as
\begin{equation}
p\left(z_k^{m,(i)}|\pi_k^{m,(i)}\right)=\left({\pi_k^{m,(i)}}\right)^{z_k^{m,(i)}}\left(1-\pi_k^{m,(i)}\right)^{1-z_k^{m,(i)}},
\label{eq:betaBernoulli}
\end{equation}
where $\pi_k^{m,(i)}$ follows beta distribution, defined by (unknown shape hyper-parameters\footnote{Notice that a beta distribution is usually defined as $p(x; \alpha, \gamma) \,\propto\, x^{\alpha -1} (1-x)^{\gamma-1}$, where $\alpha$ and $\gamma$ are two shape parameters. For simplicity we omit the dependence on $\alpha$ and $\gamma$ and keep the form as $p(x)$.}) $e_0^{(i)}$ and $f_0^{(i)}$,
\begin{equation}
p\left(\pi_k^{m,(i)}\right)=\frac{\left({\pi_k^{m,(i)}}\right)^{e_0^{(i)}-1}\left(1-\pi_k^{m,(i)}\right)^{f_0^{(i)}-1}}{\beta\left(e_0^{(i)},f_0^{(i)}\right)},\label{eq:beta}
\end{equation}
\noindent and $\beta(\cdot, \cdot)$ is the beta function. Note that in this paper we assume all the indicators are independent to each other: 
\begin{align}
    p(\bz_k^m , \bpi_k^m) 
    &= \prod_{i=1}^{n_y^m} p\left(z_k^{m,(i)}|\pi_k^{m,(i)}\right) p\left(\pi_k^{m,(i)}\right) \;,
\end{align}



The auxiliary distributions in \eqref{q_join_est}, $q(\bs_k)$, $q(\bpi_k^m)$ and $q(\bz_k^m)$, are computed by updating them sequentially and iteratively under Bayesian filtering scheme, defined as GVBKF in \cite{li2020robust}. For details of the algorithm, please refer to \cite{li2020robust} for its full derivation and Appendices \ref{Practicalaspects} for detailed practical setting up.

\section{An efficient distributed algorithm}\label{sec:distributed}

A significant limitation of the variational Kalman filter presented in the previous section is that the computation and storage requirements increase quickly with the image size, as can be attested by considering the dimension of $\bP_{k|k}$ is quadratic in the image size.
To solve this problem, we split the state $\bs_k$ into multiple groups \cite{li2023online}. The split groups are assumed to be statistically independent, following~\cite{closas2012multiple,vila2016uncertaintyMultipleKalman,vila2017multiple}.

Note that in this section, we introduce an approximation to the distributed implementation, which is part of the novelties in this paper. It is illustrated with more details starting from Eq. \eqref{Eq:statesplit}. 
To this end, we divide $\bs_k $ into $G$ groups:
\begin{align}
    \bs_k = \vect\big([\bs_k^{(1)},\ldots,\bs_k^{(G)}]\big) \,,
    \label{eq:dist_kf_groups}
\end{align}
where the elements in each block $\bs_k^{(g)}$ can be regarded as dependent, but different blocks $\bs_k^{(g_1)}$ and $\bs_k^{(g_2)}$ are assumed to be independent for $g_1\neq g_2$. Based on this assumption, the covariance matrices $\bP_{k|k-1}$ and $\bP_{k|k}$ can be defined as block diagonal matrices:
\begin{align}
    \bP_{k|k-1} &= \blkdiag\Big\{\bP_{k|k-1}^{(1)},\ldots,\bP_{k|k-1}^{(G)} \Big\} \,,
    \label{eq:dist_kf_groups_Pkold}
    \\
    \bP_{k|k} &= \blkdiag\Big\{\bP_{k|k}^{(1)},\ldots,\bP_{k|k}^{(G)} \Big\}\,.
    \label{eq:dist_kf_groups_Pk}
\end{align}
A natural question is which group structure should be selected. 
We assume a model where each block consists of all the bands of one single high-resolution pixel (resulting in $G=N_H$ blocks). Following the procedures in~\cite{closas2012multiple,vila2016uncertaintyMultipleKalman,vila2017multiple}, the distributed implementation of general VBKF under the specific independence assumption can then be derived as described in the following.





In the prediction phase of the filter, which amounts to computing the mean and covariance of $p\left(\bs_k|{\by}_{1:k-1}^{m}\right)$, according to~\cite{vila2016uncertainty} the conditional independence assumption between groups allows us to write them as:
\begin{equation}
\footnotesize
    \begin{aligned}
    \hat{\bs}_{k|k-1}^{(g)} & =\int \int \bmu_{\phi}^{(g)}\left(\bs_{k-1}^{(g)},\bs_{k-1}^{(-g)},\bu_k\right)p\left(\bs_{k-1}^{(g)}|{\by}_{1:k-1}^{m}\right) 
     \times p\left(\bs_{k-1}^{(-g)}|{\by}_{1:k-1}^{m}\right) d \bs_{k-1}^{(g)}d \bs_{k-1}^{(-g)}\,,
\label{X_predict_phase}
\end{aligned}
\end{equation}
\begin{equation}
\footnotesize
    \begin{aligned}
         \widehat{\bP}_{k|k-1}^{(g)}&=\int\int \bigg[\left(\bmu_{\phi}^{(g)}\left(\bs_{k-1}^{(g)},\bs_{k-1}^{(-g)},\bu_k\right)-\hat{\bs}_{k|k-1}^{(g)}\right)
      \left(\bmu_{\phi}^{(g)}\left(\bs_{k-1}^{(g)},\bs_{k-1}^{(-g)},\bu_k\right)-\hat{\bs}_{k|k-1}^{(g)}\right)^\top 
    \\
    & +\diag\left((\bsigma_{\phi}^2)^{(g)}\left(\bs_{k-1}^{(g)},\bs_{k-1}^{(-g)},\bu_k\right)\right) \bigg] 
     p\left(\bs_{k-1}^{(g)}|{\by}_{1:k-1}^{m}\right) p\left(\bs_{k-1}^{(-g)}|{\by}_{1:k-1}^{m}\right) d \bs_{k-1}^{(g)}d \bs_{k-1}^{(-g)}\,,
    \label{P_predic_phase}
    \end{aligned}
\end{equation}
    
where $\bmu_{\phi}^{(g)}$ and $(\bsigma_{\phi}^2)^{(g)}$ denote the subset of outputs of functions $\bmu_{\phi}$ and $\bsigma_{\phi}^2$ corresponding to the elements of the state vector in group~$g$, while the notation $\bs_{k-1}^{(-g)}$ indicates the subset of elements of vector $\bs_{k-1}$ except for those belonging to the group $g$.

These equations involve two integrals, one with respect to the variables inside the group $g$, which are expected to be important in the computation of the moments of group $g$, and another which involves the variables outside the group, which are expected to be less informative. Thus, 
we use two different integration schemes. 
In particular, the integration over $p\big(\bs_{k-1}^{(g)}|{\by}_{1:k-1}^{m}\big)$ is based on a cubature rule~\cite{arasaratnam2009cubature}, while the integration over $p\big(\bs_{k-1}^{(-g)}|{\by}_{1:k-1}^{m}\big)$ is based on random sampling.
To generate cubature samples, we have:
\begin{align}
    \bX_{i,j,k-1|k-1}^{(g)} ={} & \bL_{k-1|k-1}^{(g)} \bxi_i + \hat{\bs}_{k-1|k-1}^{(g)} \,,
\label{Eq:statesplit}
\end{align}
where $\bxi_i$ is the cubature point, for $i = 1, \dots, 2n_s$, and $\bL_{k-1|k-1}^{(g)}$ is the Cholesky decomposition of the matrix $\widehat{\bP}_{k-1|k-1}^{(g)}$~\cite{arasaratnam2009cubature}.
The random samples for variable $\bs_{k-1}^{(-g)}$ are generated following
\begin{equation}
    \Tilde{\bX}_{i,j, k-1|k-1}^{(-g)} \sim  {\mathcal{N}} \big(\hat{\bs}_{k-1|k-1}^{(-g)}, \widehat{\bP}_{k-1|k-1}^{(-g)} \big) \,,
\end{equation}
where $j = 1, \dots, \Tilde{n}_s$, with $\Tilde{n}_s$ representing the number of random samples drawn from  ${\mathcal{N}} \big(\hat{\bs}_{k-1|k-1}^{(-g)}, \widehat{\bP}_{k-1|k-1}^{(-g)}\big)$.
Following \eqref{X_predict_phase} and \eqref{P_predic_phase}, the state and covariance in the distributed form are estimated as:
\begin{align}
    \bX^*_{i,j,k|k-1} ={} & \bmu_{\phi}\left(\Tilde{\bX}_{i,j, k-1|k-1}, \bu_k\right)
    \\
    \hat{\bs}_{k|k-1}^{(g)} ={}& \frac{1}{2n_s \Tilde{n}_s} \sum_{i=1}^{2n_s} \sum_{j=1}^{\Tilde{n}_s} \bX^{*\,(g)}_{i,j,k|k-1} 
    \\
     \widehat{\bP}_{k|k-1}^{(g)} ={}& \frac{1}{2n_s \Tilde{n}_s} \sum_{i=1}^{2n_s}\sum_{j=1}^{\Tilde{n}_s} \left[\bX^{*\,(g)}_{i,j,k|k-1} - \hat{\bs}_{k|k-1}^{(g)} \right] \times \left[\bX^{*\,(g)}_{i,j,k|k-1} - \hat{\bs}_{k|k-1}^{(g)} \right]^\top 
     \nonumber\\
    & + (\bsigma_{\phi}^2)^{(g)}\left(\Tilde{\bX}_{i,j, k-1|k-1},\bu_k\right) \,,
\end{align}
where $\Tilde{\bX}_{i,j, k-1|k-1}$ is the vector whose elements inside and outside group $g$ are equal to those of $\bX_{i,j,k-1|k-1}^{(g)}$ and $\bX_{i,j,k-1|k-1}^{(-g)}$, respectively.

Since the proposed distributed general VBKF is based on the Kalman filter scheme, the distributed implementation in terms of measurement update phase is quite similar to the one in \cite{li2023online}, while the difference appears to the Kalman filter gain and 
a posterior covariance, caused by the robust definition of observation covariance. The update equations are given by:
\begin{equation}
    \hat{\bs}_{k|k}^{(g)} = \hat{\bs}_{k|k-1}^{(g)}  + \bK_{k}^{(g)} \left({\by}_{k}^{m}-{\bH}^{m}\hat{\bs}_{k|k-1}\right) \,,
\end{equation}
\begin{equation}
\widehat{\bP}_{k|k}^{(g)}=\widehat{\bP}_{k|k-1}^{(g)}-\bK_k^{(g)}\left(\bS_k+\langle \bSigma_k^{-1}(\bz_k^m) \rangle^{-1}\right
)\big(\bK_k^{(g)}\big)^\top\,,
\label{P_update_split}
\end{equation}
\begin{equation}
\bK_{k}^{(g)}=\bC_k^{(g)}\left(\bS_{k}+\langle \bSigma_k^{-1}(\bz_k^m) \rangle^{-1}\right)^{-1}\,,
\label{K_update_split}
\end{equation}
where
\begin{align}
\bC_k^{(g)} &=\int \left(\bs_k^{(g)}-\hat{\bs}_{k|k-1}^{(g)}\right)\left({\bH}^{m}\bs_k-{\bH}^{m}\hat{\bs}_{k|k-1}\right)^\top   
p\left(\bs_k|{\by}_{1:k-1}^{m}\right) d\bs_k \,,
\label{CS_t1}
\\
\bS_k &=\int \left({\bH}^{m}\bs_k-{\bH}^{m}\hat{\bs}_{k|k-1}\right)\left({\bH}^{m}\bs_k-{\bH}^{m}\hat{\bs}_{k|k-1}\right)^\top  
p\left(\bs_k|{\by}_{1:k-1}^{m}\right) d\bs_k \,,
\label{S_t1}
\end{align}
and $\bz_k^m$ represents one of the $2^{n_y^m}$ possible combinations of $\{z_k^{m,(i)}\}_{i=1}^{n_y^m}$ binary values; 
the set of all possible combinations is given by $\mathcal{Z}=\{0,1\}^{n_y^m}$ such that $|\mathcal{Z}|=2^{n_y}$. 
The expectation of $ \bSigma_k^{-1}(\bz_k^m)$ with respect to $q(\bz_k^m)$ is defined as 
\begin{equation}\label{eq:expzSigmaInv}
    \langle \bSigma_k^{-1}(\bz_k^m) \rangle=\sum\limits_{\bz\in\mathcal{Z}} \bSigma_k^{-1}(\bz) \; q(\bz_k^m=\bz) \;.
\end{equation}
\section{Experiments and Results}
\label{sec:experiments}
In this section, we use the proposed algorithm to fuse Landsat-8 and MODIS image sequences to generate images with high spatial and temporal resolutions. We compare the proposed approach, called GIVKF-NN, to three competing recursive algorithms, illustrating the importance of both the proposed outlier and temporal evolution models.
The first is a linear Kalman filter with a simple data-driven model for the dynamical image evolution proposed in~\cite{li2023online}, which we refer to as KF.
The second is an extension of a Kalman filter with the learned temporal evolution model proposed in Section~\ref{sec:NN} which we refer to as KF-NN.
The third is the distributed variational Bayesian fusion framework proposed in Section~\ref{sec:VBKF} but using the simple dynamical evolution model of~\cite{li2023online} instead of the proposed learned dynamical model, which we refer to as GIVKF.
In the following, we describe the data and experiment setup, and analyze the experiment results.
 

\subsection{Study region}
We consider two sites and two settings (cloudless and cloudy cases).
The first is the Oroville dam. It is located on the Feather River, in the Sierra Nevada Foothills (38° 35.3' N and 122° 27.8' W), as the tallest dam in USA. It acts as a major water storage facility in California State Water Project, with a maximum storage capacity of $1.54\times10^{11}$ ft$^{3}$.

We focus on two particular areas of the Oroville dam delimited by the blue and orange boxes in the left panel of Figure~\ref{fig:site-location}, for cloudy and cloudless cases, respectively.

The other site is the Elephant Butte reservoir (Figure~\ref{fig:site-location}, right panel). Located in the southern part of the Rio Grande river, in NM, USA (33° 19.4' N and 107° 26.2' W), it is the largest reservoir in New Mexico, providing power and irrigation to southern New Mexico and Texas. 
It is at an elevation of 4,414 ft,
and has a surface area of 36,500 acres.
We focus at a particular area of the Elephant Butte reservoir delimited by the blue box in the lower panel of Figure~\ref{fig:site-location}, for both cloudy and cloudless cases.
\begin{figure}
    \centering
    \includegraphics[width=0.7\linewidth]{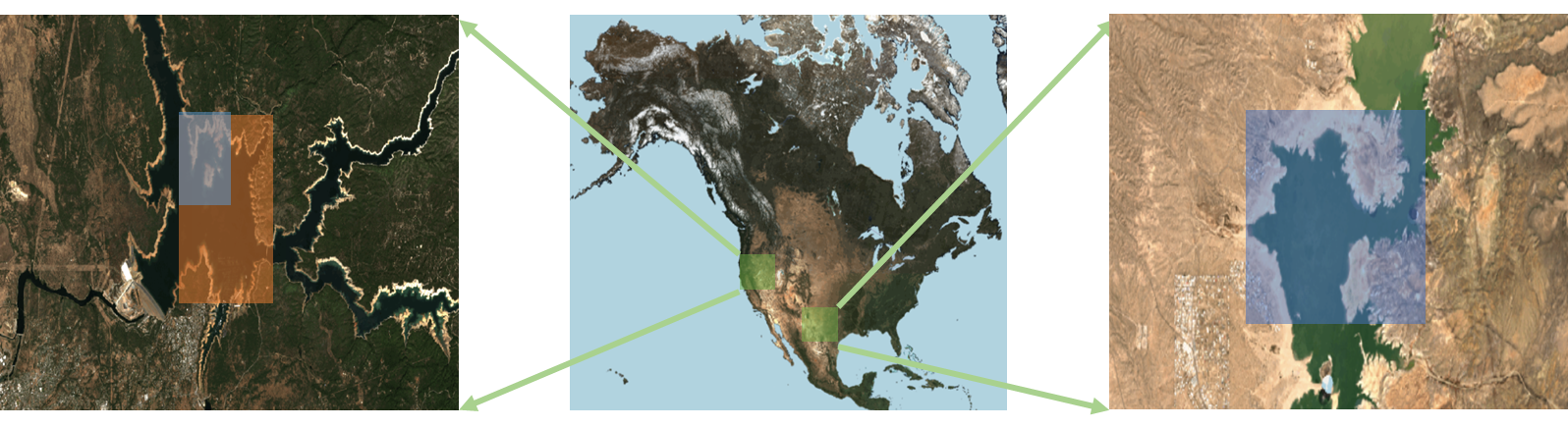}\\
    \vspace{-0.3cm}
    \caption{The left panel shows the Oroville dam site. The blue and orange boxes delimit the study area used in the cloudy and cloudless datasets, respectively. The right panel shows the Elephant Butte site. The blue box delimits the specific study area used in the cloudy and cloudless datasets.}
    \label{fig:site-location}
\end{figure}
\subsection{Remotely sensed data}


For the Elephant Butte site, we evaluate the performance of the algorithms when processing a larger geographical area of approximately $9km \times 9km$, corresponding to Landsat and MODIS images with $324\times 324$ and $36 \times 36$ pixels, respectively.
We focus on the red and near-infrared (NIR) bands of the Landsat and MODIS instruments since they are often used to distinguish water from other land cover elements in the image~\cite{gao1996ndwi}. To train and test the NNs and set up the algorithm, we collect 47 Landsat images as the  historical dataset~$\cp{D}$, with observed dates varying from 2014/01/16 to 2017/11/24.


In the cloudless case, after removing images with significant cloud cover between dates {2019/03/19 and 2019/07/09}, we obtained a set of 5 Landsat and 6 MODIS images to process. The first Landsat image is used to initialize the algorithms, while two others acquired at dates 06/07 and 06/23 were set aside as ground truth to measure the performance of different algorithms.
Therefore, the remaining 6 MODIS and 2 Landsat images are processed by the algorithms, and we evaluate their performance through the fused image results at dates 06/14 and 06/27 (where the MODIS observations were available), since the MODIS images at dates 06/07 and 06/23 (matching the ground truth) had to be discarded due to significant cloud cover.
As for the cloudy case, the Landsat data is the same as for the cloudless case, while for the MODIS data we substitute a clean image at date 06/27 by another one acquired at 06/19, which is contaminated by a cloud.

For the experiments with the Oroville Dam site, we focus on a smaller area exploring the performance of proposed algorithms under a more extreme case, with clouds covering the whole observed area.
We also evaluate the methods on the cloudless dataset of the Oroville Dam site for comparison.
The study region, shown in the left panel of Figure~\ref{fig:site-location}, corresponds to Landsat and MODIS images with $81\times 81$ and $9\times 9$ pixels for cloudy case (marked with an orange square), and  $162\times 162$ and $18\times 18$ pixels for cloudless case (marked with a blue square), respectively.
We collected 50 Landsat data from 2013/04/09 to 2017/12/07 to serve as the historical dataset~$\cp{D}$, used for training the NNs and setting up the algorithms. In this experiment we will also focus on the red and NIR bands.

For the cloudless case, we collected MODIS and Landsat data acquired on the interval ranging from 2018/07/03 to 2018/09/21.
After filtering for heavy cloud cover, a set of 6 Landsat and 15 MODIS images were obtained. We used the first Landsat image to initialize the algorithms. For the remaining 5 Landsat images, we have 3 of them, acquired at 07/19, 08/20 and 09/05, not processed by the algorithm but only use as ground truth, while the corresponding MODIS images are used to estimate the fused image for comparison.

For the cloudy case, we collected MODIS and Landsat data acquired 
from 2018/03/29 to 2018/07/19. 
We filtered the images in this interval for heavy cloud cover, with the exception of the date 2018/05/16, for which the corresponding Landsat image is contaminated by heavy cloud cover. This led to 6 Landsat and 10 MODIS images. The first Landsat image is used as initialization for the algorithms, and 2 out of the remaining 5 Landsat images (acquired at dates 06/01 and 07/03) were set aside for evaluation and not processed by the algorithms. They were used to evaluate the images reconstructed by the algorithms from the MODIS observations at the same dates.







\subsection{Algorithm setup}

We initialized all algorithms using a high resolution Landsat observation, i.e., $\bs_{0|0}={\by}_0^{\mathsf{L}}$, and set all the hyperparameters according to experiments performed over the historical dataset $\calD$ and statistics computed from it.
From several tests we set the parameters of algorithm as shown in the following.


We set $\bP_{0|0}=10^{-10}\bP_0$, where $\bP_0 = \frac{1}{10}\mathbb{1}+\frac{9}{10}\bI $ for Oroville Dam and $\bP_0 = \frac{1}{2}\mathbb{1}+\frac{1}{2}\bI$ for Elephant Butte, with $\mathbb{1}$ being an all-ones matrix.  In Oroville Dam, the noise covariance matrices were set as $\bR^{\mathsf{L}} = 3\times 10^{-2} \bP_1^{\mathsf{L}}$ and $\bR^{\mathsf{M}}=10^{-4}\bP_1^{\mathsf{M}}$, where
$\bP_1^m = \bI_{m} \otimes \begin{bmatrix}
    1 & 0.1 \\
    0.1 & 2
\end{bmatrix}$, with $\otimes$ denoting the Kronecker product,  $\bI_{m}$ represents the identity matrix, and the superscript $m\in\{{\mathsf{M}},{\mathsf{L}}\}$  represents the adjustment of the corresponding covariance size for MODIS and Landsat images.
In the Elephant Butte dataset, the noise covariance matrices were set as $\bR^{\mathsf{L}} = 7.5\times 10^{-3} \bP_0^{\mathsf{L}}$ and $\bR^{\mathsf{M}}=2.5 \times 10^{-5}\bP_0^{\mathsf{M}}$, 
where the superscript $m\in\{{\mathsf{M}},{\mathsf{L}}\}$ of $\bP_0^m$ represents the adjustment of the corresponding covariance size for MODIS and Landsat images.
%
The blurring and downsampling matrices were set as $\bH^{\mathsf{L}}=\bI$ for Landsat, while for MODIS $\bH^{\mathsf{M}}$ is defined as a spatial convolution based on a uniform $9\times9$ filter, defined by $\bh=\frac{1}{81}{\mathbb{1}_{9\times9}}$ (where ${\mathbb{1}_{9\times9}}$ is a $9\times9$ matrix of ones), followed by decimation by a factor of~$9$ and a bandwise gain factor to compensate scaling differences between Landsat and MODIS sensors. This represents the degradation occurring at the sensor (see, e.g.,~\cite{huang2013spatiotemporalFusionBayes}).


The hyper-parameter of GVIKF represents the prior knowledge of the outliers information, which is different between the two datasets. Histograms are shown in Appendix~\ref{app:SupportiveExperimentsResults} to illustrate the difference in the outlier magnitudes between the two datasets. In this paper, $e_0 = 0.98$ and $f_0 = 0.02$ in Elephant Butte, while $e_0 = 0.5$ and $f_0 = 0.5$ in Oroville Dam. 
To limit the computational cost, the total number of random samples is set as as $\tilde{n}_s=8$ for each $\bs_{k}^{(-g)}$ in the distributed implementation.
Further details on the setup of the proposed algorithm can be found in Appendix~\ref{Practicalaspects}.


To measure the performance of the methods, two metrics are introduced in this paper, the Root Mean Square Error (RMSE) and Misclassification Percentage (MP), defined as
    $\text{RMSE}_k = \sqrt{\frac{1}{L_H N_H} \big\|\bs_k - \hat{\bs}_{k|k}\big\|_2^2}$, 
    $\text{MP}_k = \frac{100 \%}{N_H} \|\be_k - \hat{\be}_{k|k}\|_1$, 
where $\bs_k$ represents the reference (ground truth) image with $L_H$ bands and $N_H$ pixels,
$\hat{\bs}_{k|k}$ represents the images estimated by the various methods, and $\be_k$ and $\hat{\be}_{k|k}$ represent binary classification results generated by the K-means algorithm applied to $\bs_k$ and $\hat{\bs}_{k|k}$, respectively.
\subsection{Results and discussion for the Elephant Butte site}
We fused the red and NIR bands of MODIS and Landsat for the Elephant Butte site in both cloudless and cloudy scenarios. Note that we only show the NIR reflectance band for the reconstruction result for succinctness. The results for the red band are included in Appendix \ref{app:SupportiveExperimentsResults}.
In the cloudless scenario, the fusion results for NIR band and all algorithms are shown in Figure~\ref{fig:Reconstruction-EButte-cloudless} along with the acquired MODIS and Landsat images. In the top labels, acquisition dates represent the acquired date. If the images are processed by the fusion algorithm, we have $M$ for MODIS and $L$ for Landsat, following the date in top label. As shown in the figure, only the first and last Landsat images were used in the fusion process, while the remaining two images at dates 06/07 and 06/23 act as ground-truth. They are used to evaluate the accuracy of fused images at 06/14 and 06/27 to compare the performances of different methods (note that the MODIS images at dates 06/07 and 06/23 were not available due to cloud cover).
Figure~\ref{fig:watermaperror-Ebutte-cloudless} shows the misclassification maps (i.e., the absolute error between the water maps by different algorithm and the ground-truth). It can be seen that all four methods share visually similar image reconstructions.

To evaluate the performances of different methods more clearly, Table~\ref{Table:misclassification-rmse-ebutte-cloudless} shows the misclassification percentage and the RMSE of the estimated images. In terms of RMSE, the GVIKF-NN and KF-NN hold a close performance on average. GVIKF and KF, on the other hand, show a higher RMSE.
In terms of misclassification percentage, we can see that the GVIKF-NN holds the smallest misclassification percentage on average, and the second best method is the KF-NN, followed by KF and GVIKF. This shows that the NN model is important for preventing performance losses in the proposed robust framework in the absence of outliers.

In the cloudy scenario, the fusion results for the NIR band and all algorithms are shown in Figure~\ref{fig:Reconstruction-EButte-cloudy}, while Figure~\ref{fig:watermaperror-Ebutte-cloudy} shows the misclassification mappings. To compare different methods with the proposed one, the Landsat images at dates 06/07 and 06/23 act as ground truth to measure the accuracy of fused images at dates 06/14 and 06/27 (the MODIS images at dates 06/07 and 06/23 were not available due to cloud cover). Note that in the MODIS at 06/19, part of the pixels in the top part are covered by a cloud. 
It can be seen that the KF is heavily influenced by the cloudy image, followed by GVIKF. The cloud in the images estimated by those methods at 06/19 is basically not removed at all. The proposed GIVKF-NN, on the other hand, is able to suppress the outlier, while KF-NN shows intermediate performance.
This indicates that the robustness of the outlier-aware variational inference framework against cloudy pixels depends on the accuracy of the dynamical image evolution model, and shows that considering both the robust methodology and the learned NN evolution model is key to obtain a high performance in GIVKF-NN. 

The quantitative results are shown in Table~\ref{Table:misclassification-rmse-ebutte-cloudy}, which contains the misclassification percentage and the RMSE of estimated images of the different methods. The GVIKF-NN and KF-NN hold the first and second best performance on average in terms of both misclassification percentage and RMSE.
The GVIKF and the KF, which do not have an accurate dynamical evolution model, show significantly worse performance.
In summary, the results in Elephant Butte have showed the learned NN-based model, by representing the image evolution more accurately, brings important improvements to both the GVIKF and KF methods. 
\begin{table}[ht]
\centering
\caption{Misclassification and RMSE performances for the Elephant Butte site in the cloudless scenario.}
\scriptsize
\vspace{-0.15cm}
\begin{tabular}{|ccccc|}
\hline
\multicolumn{1}{|c|}{}             & \multicolumn{1}{c|}{KF}     & \multicolumn{1}{c|}{KF-NN}  & \multicolumn{1}{c|}{GVIKF}  & GVIKF-NN \\ \hline
\multicolumn{5}{|c|}{Misclassification Percentage $\%$}                                                                                     \\ \hline
\multicolumn{1}{|c|}{06/14(06/07)} & \multicolumn{1}{c|}{6.1204} & \multicolumn{1}{c|}{5.2231} & \multicolumn{1}{c|}{6.1338} & 5.1917   \\
\multicolumn{1}{|c|}{06/27(06/23)} & \multicolumn{1}{c|}{7.1397} & \multicolumn{1}{c|}{5.3612} & \multicolumn{1}{c|}{7.2912} & 5.2879   \\
\multicolumn{1}{|c|}{07/09}        & \multicolumn{1}{c|}{7.1007} & \multicolumn{1}{c|}{5.5860} & \multicolumn{1}{c|}{7.2140} & 5.5794   \\ \hline
\multicolumn{1}{|c|}{Average}      & \multicolumn{1}{c|}{6.7869} & \multicolumn{1}{c|}{5.3901} & \multicolumn{1}{c|}{6.8797} & 5.3530   \\ \hline
\multicolumn{5}{|c|}{RMSE}                                                                                                              \\ \hline
\multicolumn{1}{|c|}{06/14(06/07)} & \multicolumn{1}{c|}{0.0046} & \multicolumn{1}{c|}{0.0034} & \multicolumn{1}{c|}{0.0046} & 0.0034   \\
\multicolumn{1}{|c|}{06/27(06/23)} & \multicolumn{1}{c|}{0.0050} & \multicolumn{1}{c|}{0.0032} & \multicolumn{1}{c|}{0.0052} & 0.0031   \\
\multicolumn{1}{|c|}{07/09}        & \multicolumn{1}{c|}{0.0050} & \multicolumn{1}{c|}{0.0038} & \multicolumn{1}{c|}{0.0050} & 0.0038   \\ \hline
\multicolumn{1}{|c|}{Average}      & \multicolumn{1}{c|}{0.0049} & \multicolumn{1}{c|}{0.0035} & \multicolumn{1}{c|}{0.0049} & 0.0035   \\ \hline
\end{tabular}
\label{Table:misclassification-rmse-ebutte-cloudless}
\end{table}
\begin{table}[ht]
\centering
\caption{Misclassification and RMSE performance for the Elephant Butte site in the cloudy scenario.}
\scriptsize
\vspace{-0.4cm}
\begin{tabular}{|c|c|c|c|c|}
\hline
        & \multicolumn{1}{c|}{KF} & \multicolumn{1}{c|}{KF-NN} & \multicolumn{1}{c|}{GVIKF} & \multicolumn{1}{c|}{GVIKF-NN} \\ \hline
\multicolumn{5}{|c|}{Misclassification Percentage $\%$}  \\
\hline
06/14(06/07) & 6.1204                  & 5.2231                     & 6.1043                     & 5.1907                        \\ 
06/19(06/23) & 11.3521                 & 6.1719                     & 11.3588                    & 5.7318                        \\ 
07/09        & 7.4703                  & 5.6899                     & 7.5332                     & 5.8023                        \\ \hline
Average      & 8.3143                  & 5.6950                     & 8.3321                     & 5.5749                        \\ \hline

\multicolumn{5}{|c|}{RMSE}  \\
\hline

06/14(06/07) & 0.0046                  & 0.0034                     & 0.0046                     & 0.0034                        \\ 
06/19(06/23) & 0.0132                  & 0.0054                     & 0.0129                     & 0.0046                        \\ 
07/09        & 0.0060                  & 0.0045                     & 0.0059                     & 0.0046                        \\ \hline
Average      & 0.0079                  & 0.0045                     & 0.0078                     & 0.0042                        \\ \hline
\end{tabular}
\label{Table:misclassification-rmse-ebutte-cloudy}
\end{table}
\begin{figure*}[h!]
  \centering
  \includegraphics[width=0.95\linewidth, height = 0.416\linewidth]{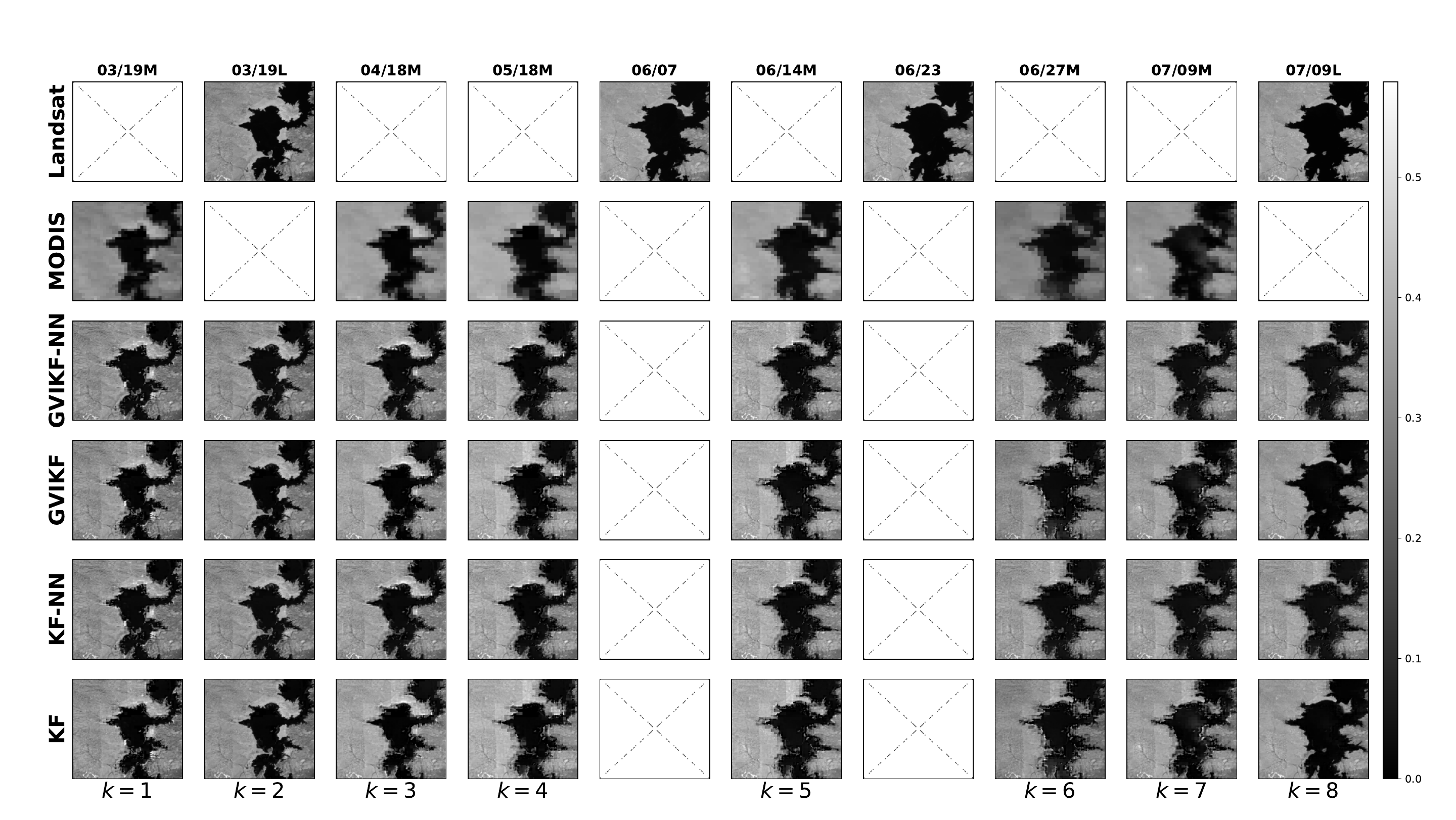}
  \label{fig:Reconstruction2-EButte-cloudless}
  \vspace{-1.5ex}
  \caption{Fusion results for the NIR band from MODIS (band 2) and Landsat (band 5) in the Elephant Butte in cloudless case. The first two rows show the observed images at MODIS and Landsat bands. At each time index the fused images by GVIKF-NN, GVIKF, KF-NN and KF are presented at the following rows. Note that some Landsat images were used solely as ground-truth with only the acquisition date as the top label, while images processed by the algorithms are indicated on top labels where ``M'' stands for MODIS and ``L'' for Landsat.}\label{fig:Reconstruction-EButte-cloudless}
\end{figure*}
\begin{figure}[h]
  \centering
  \centering
  \includegraphics[width=0.7\linewidth]{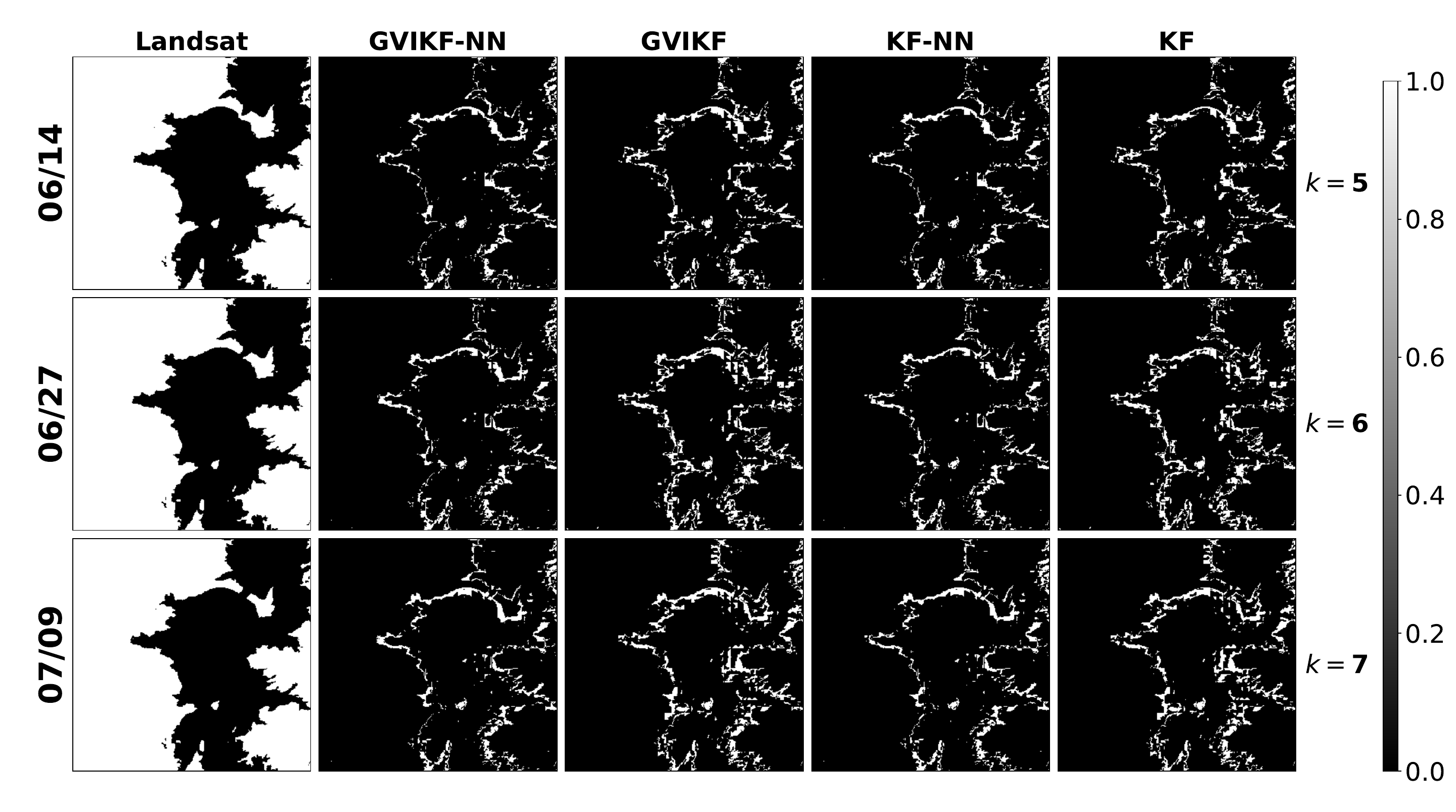}
\vspace{-0.5cm}
\caption{Absolute error of water map of images for the Elephant Butte example in cloudless case based on K-means clustering strategy. In this plot, 0 pixel value indicates correct classification and 1 pixel value indicates misclassification. For comparison, acquired Landsat images are shown in the first column as the ground-truth.} 
\label{fig:watermaperror-Ebutte-cloudless}
\end{figure}
\begin{figure*}[h!]
  \centering
  \includegraphics[width=0.95\linewidth, height = 0.416\linewidth]{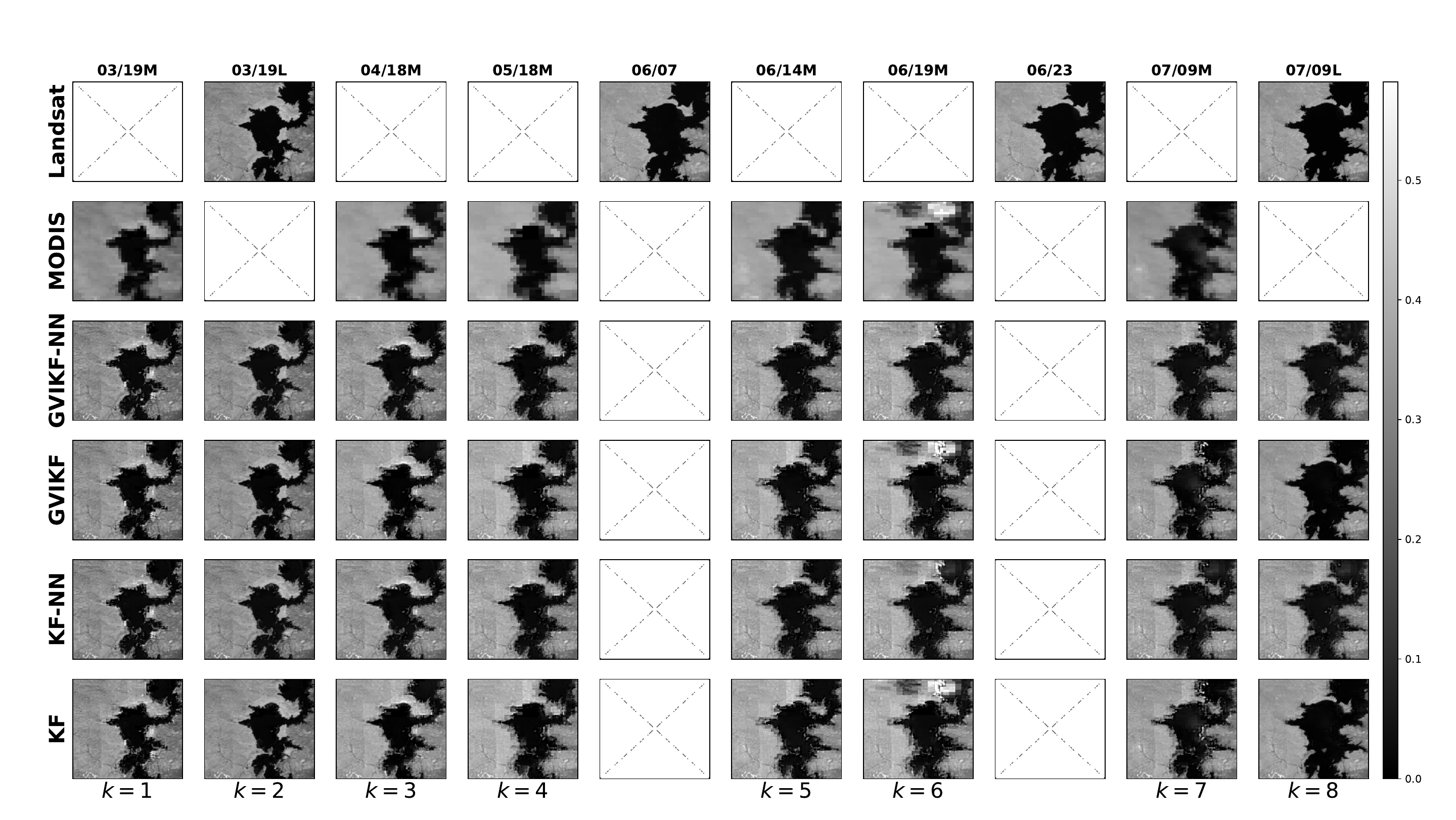}
  \label{fig:Reconstruction2-EButte-cloudy}
  \vspace{-1.5ex}
  \caption{Fusion results for the NIR band from MODIS (band 2) and Landsat (band 5) for the Elephant Butte example in cloudy case. The first two rows show the observed images at MODIS and Landsat bands. At each time index the fused images by GVIKF-NN, GVIKF, KF-NN and KF are presented at the following rows. Note that some Landsat images were used solely as ground-truth with only acquisition date as top labels, while images processed by the algorithms are indicated on top labels where ``M'' stands for MODIS and ``L'' for Landsat.}\label{fig:Reconstruction-EButte-cloudy}
\end{figure*}
\begin{figure}[h]
  \centering
  \centering
    \includegraphics[width=0.7\linewidth]{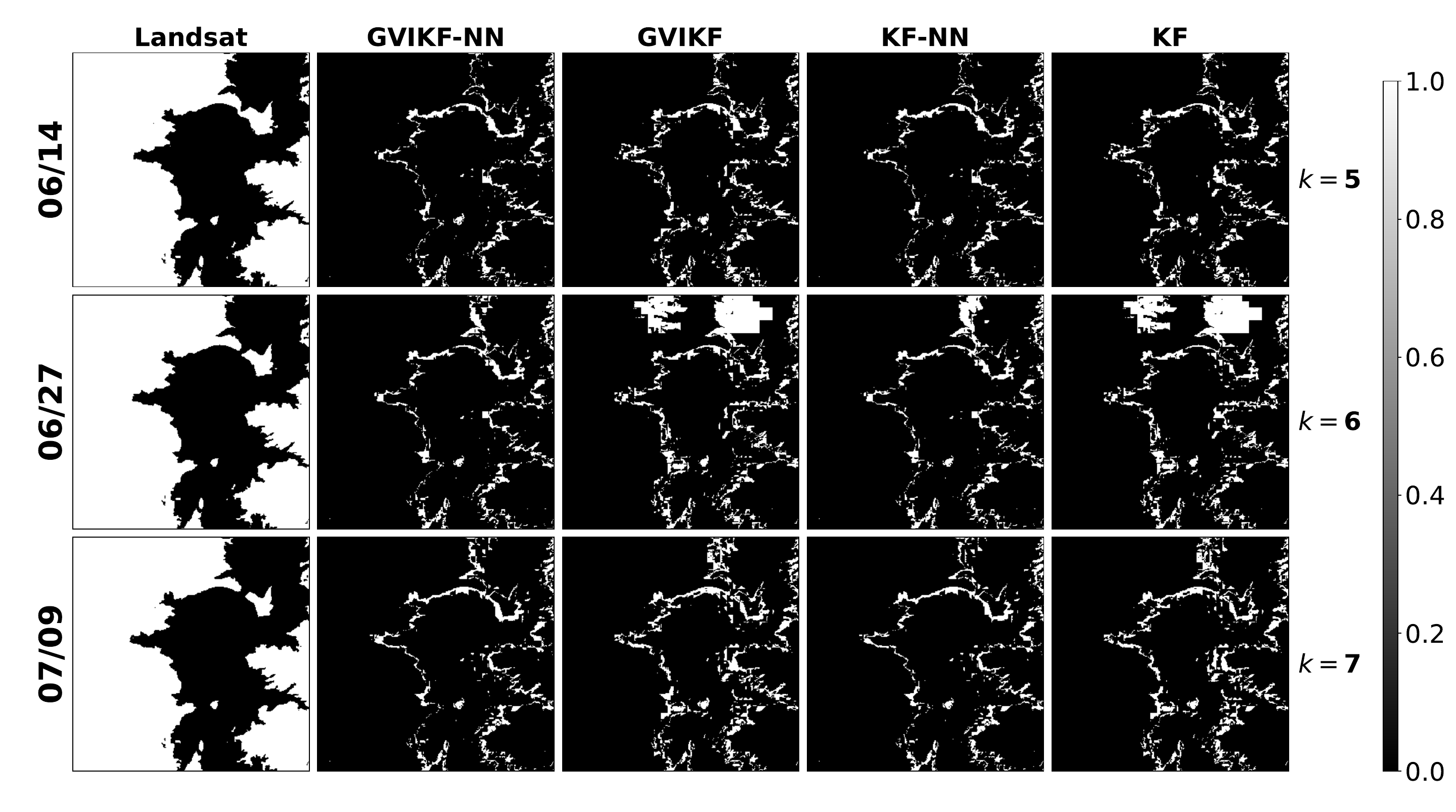}
\vspace{-0.5cm}
\caption{Absolute error of water map of images for the Elephant Butte example in cloudy case based on K-means clustering strategy. In this plot, 0-valued pixels indicate correct classification and 1-valued pixels indicate misclassification. For comparison, acquired Landsat images are shown in the first column as the ground-truth.} 
\label{fig:watermaperror-Ebutte-cloudy}
\end{figure}
\subsection{Results and discussion for the Oroville Dam site}
\begin{table}[ht]
\centering
\caption{Misclassification and RMSE performance for the Oroville Dam site in the cloudless scenario.}
\scriptsize
\vspace{-0.4cm}
\begin{tabular}{|ccccc|}
\hline
\multicolumn{1}{|c|}{}        & \multicolumn{1}{c|}{KF}      & \multicolumn{1}{c|}{KF-NN}  & \multicolumn{1}{c|}{GVIKF}  & GVIKF-NN \\ \hline
\multicolumn{5}{|c|}{Misclassification Percentage $\%$}                                                                                 \\ \hline
\multicolumn{1}{|c|}{07/19}   & \multicolumn{1}{c|}{5.8528}  & \multicolumn{1}{c|}{4.0390} & \multicolumn{1}{c|}{4.0047} & 3.4903   \\
\multicolumn{1}{|c|}{08/20}   & \multicolumn{1}{c|}{8.6267}  & \multicolumn{1}{c|}{7.4608} & \multicolumn{1}{c|}{7.9904} & 7.5636   \\
\multicolumn{1}{|c|}{09/05}   & \multicolumn{1}{c|}{10.2157} & \multicolumn{1}{c|}{8.4438} & \multicolumn{1}{c|}{9.5946} & 8.8820   \\
\multicolumn{1}{|c|}{09/21}   & \multicolumn{1}{c|}{10.4557} & \multicolumn{1}{c|}{9.3698} & \multicolumn{1}{c|}{9.7470} & 9.7813   \\ \hline
\multicolumn{1}{|c|}{Average} & \multicolumn{1}{c|}{8.7877}  & \multicolumn{1}{c|}{7.3283} & \multicolumn{1}{c|}{7.8342} & 7.4293   \\ \hline
\multicolumn{5}{|c|}{RMSE}                                                                                                          \\ \hline
\multicolumn{1}{|c|}{07/19}   & \multicolumn{1}{c|}{0.0029}  & \multicolumn{1}{c|}{0.0019} & \multicolumn{1}{c|}{0.0021} & 0.0016   \\
\multicolumn{1}{|c|}{08/20}   & \multicolumn{1}{c|}{0.0054}  & \multicolumn{1}{c|}{0.0041} & \multicolumn{1}{c|}{0.0047} & 0.0038   \\
\multicolumn{1}{|c|}{09/05}   & \multicolumn{1}{c|}{0.0050}  & \multicolumn{1}{c|}{0.0033} & \multicolumn{1}{c|}{0.0045} & 0.0034   \\
\multicolumn{1}{|c|}{09/21}   & \multicolumn{1}{c|}{0.0056}  & \multicolumn{1}{c|}{0.0040} & \multicolumn{1}{c|}{0.0050} & 0.0040   \\ \hline
\multicolumn{1}{|c|}{Average} & \multicolumn{1}{c|}{0.0047}  & \multicolumn{1}{c|}{0.0033} & \multicolumn{1}{c|}{0.0041} & 0.0032   \\ \hline
\end{tabular}
\label{Table:misclassification-rmse-oroville-cloudless}
\vspace{-0.4cm}
\end{table}
\begin{table}[ht]
\centering
\caption{Misclassification and RMSE performance for the Oroville Dam site in the cloudy scenario.}
\scriptsize
\vspace{-0.4cm}
\begin{tabular}{|ccccc|}
\hline
\multicolumn{1}{|c|}{}        & \multicolumn{1}{c|}{KF}      & \multicolumn{1}{c|}{KF-NN}   & \multicolumn{1}{c|}{GVIKF}   & GVIKF-NN \\ \hline
\multicolumn{5}{|c|}{Misclassification Percentage $\%$}                                                                                   \\ \hline
\multicolumn{1}{|c|}{05/16}   & \multicolumn{1}{c|}{43.3318} & \multicolumn{1}{c|}{16.2780} & \multicolumn{1}{c|}{10.9739} & 8.2000   \\
\multicolumn{1}{|c|}{06/01}   & \multicolumn{1}{c|}{10.8215} & \multicolumn{1}{c|}{8.0323}  & \multicolumn{1}{c|}{8.5048}  & 8.9468   \\
\multicolumn{1}{|c|}{07/03}   & \multicolumn{1}{c|}{8.2000}  & \multicolumn{1}{c|}{5.6851}  & \multicolumn{1}{c|}{7.4684}  & 4.1000   \\
\multicolumn{1}{|c|}{07/19}   & \multicolumn{1}{c|}{8.7334}  & \multicolumn{1}{c|}{7.8342}  & \multicolumn{1}{c|}{7.9866}  & 2.5911   \\ \hline
\multicolumn{1}{|c|}{Average} & \multicolumn{1}{c|}{17.7717} & \multicolumn{1}{c|}{9.4574}  & \multicolumn{1}{c|}{8.7334}  & 5.9595   \\ \hline
\multicolumn{5}{|c|}{RMSE}                                                                                                            \\ \hline
\multicolumn{1}{|c|}{05/16}   & \multicolumn{1}{c|}{0.1029}  & \multicolumn{1}{c|}{0.0162}  & \multicolumn{1}{c|}{0.0132}  & 0.0072   \\
\multicolumn{1}{|c|}{06/01}   & \multicolumn{1}{c|}{0.0080}  & \multicolumn{1}{c|}{0.0081}  & \multicolumn{1}{c|}{0.0085}  & 0.0075   \\
\multicolumn{1}{|c|}{07/03}   & \multicolumn{1}{c|}{0.0053}  & \multicolumn{1}{c|}{0.0044}  & \multicolumn{1}{c|}{0.0078}  & 0.0039   \\
\multicolumn{1}{|c|}{07/19}   & \multicolumn{1}{c|}{0.0069}  & \multicolumn{1}{c|}{0.0064}  & \multicolumn{1}{c|}{0.0093}  & 0.0046   \\ \hline
\multicolumn{1}{|c|}{Average} & \multicolumn{1}{c|}{0.0308}  & \multicolumn{1}{c|}{0.0088}  & \multicolumn{1}{c|}{0.0097}  & 0.0058   \\ \hline
\end{tabular}
\label{Table:misclassification-rmse-oroville-cloudy}
\vspace{-0.4cm}
\end{table}

We now compare the four algorithms on the Oroville Dam example with and without clouds. The setup is similar to the one for the Elephant Butte example. Considering space limitations, only quantitative results are provided. Please refer to the supplemental material for the reflectance and water mapping results.
For the cloudless scenario, Table~\ref{Table:misclassification-rmse-oroville-cloudless} shows the misclassification percentage and the RMSE of the images estimated by the different methods when they were compared to the ground-truth. It can be seen that the GVIKF-NN holds the smallest RMSE and misclassification percentage on average, while the second best method is the KF-NN, followed by GVIKF and KF. Like in the Elephant Butte example, this illustrates the performance improvements brought by the use of the proposed temporal image evolution model based on NNs when no outliers are present.


We also evaluate the methods in the cloudy scenario, when the Landsat image at date 05/16 is completely covered by a large cloud present over the observed area, constituting a more extreme case of outlier contamination.
Note that since there is no cloudless Landsat image around date 05/16 available for use as ground-truth, we used interpolation method to estimate a surrogate ground-truth of the Landsat image at 05/16 from the Landsat observations at dates 04/14 and 06/01. 
The quantitative results are shown in Table~\ref{Table:misclassification-rmse-oroville-cloudy}, where the results at 05/16 are based on the ground-truth obtained by the interpolation method.
It can be seen that the results estimated by KF are heavily influenced by the large cloud cover in the Landsat observation at 05/16. On the other hand, the GVIKF, KF-NN and GVIFK-NN methods hold a comparatively stable performance, keeping estimated images robust against large cloud cover. Specifically, GVIKF-NN outperforms the other methods on average, followed by the GIVKF and KF-NN, which showed similar performance. The KF had the worst performance by far.
This illustrates the advantages of combining both the robust framework and the learned NN-based model. Specifically, in the presence of extreme amounts of outliers in the image the robust methodology in GIVKF is able to provide a more competitive performance and mitigate the presence of the outliers with the KF-NN even without an accurate evolution model.
\section{Conclusion}
\label{sec:conclusion}
In this paper, we proposed a recursive image fusion method based on location-aware NNs that is robust to outliers such as clouds and shadows. To achieve this goal we proposed an imaging model where the acquisitions are contaminated by discrete outliers. The stochastic time evolution of the high-resolution images is represented by a NN learned from a small set of historical images. To estimate the high spatial-temporal-resolution image sequence, we resorted to a variational Bayesian filtering framework. A distributed approximate solution that is scalable to large datasets was also proposed.
Experimental results show that the proposed algorithm accounting for both outliers and learned dynamical model is more robust against cloud cover without losing performance when no clouds are present.
\section{Acknowledgements}
This work was supported by the French National Research Agency, under grants ANR-23-CE23-0024, ANR-23-CE94-0001, and by the National Science Foundation, under Awards NSF2316420, ECCS-1845833 and CCF-2326559. 
\setcounter{section}{0} 
\section*{Supplementary material to 'Robust Recursive Fusion of Multiresolution Multispectral Images with Location-Aware Neural Networks' by H. Li, R. Borsoi, T. Imbiriba, P. Closas}

\section{Model structure and parameters}
\label{Practicalaspects}
\subsection{Practical aspects of GVBKF application}

The practical aspects of algorithm considered in this paper is the same as in \cite{li2020robust}. 
In this paper, we have two criteria. The first one is a threshold, measuring the change between two consecutive state estimates during iterations ($10\%$ relative difference in the experiments of this paper), while the second one is the maximum of the iteration amount ($20$ iterations in the experiments of this paper).

\subsection{NN structure and parameters}
\label{app:CNN_architecture}

This subsection aims to show detailed structure and parameters used in NN and its cost function.
In terms of  ${\rm NN}_{\phi}^Q$ in~\eqref{eq:NN_model_sigma}, we selected its architecture as 
\[{\rm NN}_{\phi}^Q = \text{Conv1d}_1(\text{flatten}(\text{Conv2d}_2(\text{Conv2d}_1(\mathtt{input})))))\]
where $\mathtt{input}$ contains $\big[\bs_{k-1},\, \boldsymbol{pos}^\top,\, \bq_0^\top,\, date_k \big]^\top$ ordered as a tensor with spatial size $81 \times 81$ and $7$ channels, where $\bs_{k-1}$ containing $2$ bands covers 2 channels, $\boldsymbol{pos}$ covers 2 channels as it contains the $(x,y)$ relative spatial position of each pixel with respect to the top-left corner, $\bq_0$, representing the diagonal elements of the covariance of $\bs_k$ in the historical dataset for each band, covers 2 channels, and $date_k$ covers the final channel. $\text{Conv2d}_1$ has $7$ \textit{inchannels} (corresponding to the $7$ input channels) and $12$ \textit{outchannels}, with \textit{kernelsize} $= 9$, \textit{stride} $= 1$, \textit{padding} $= 4$, \textit{paddingmode} as 'zero', \textit{dilation} $= 1$, \textit{groups} $= 4$, \textit{bias} as 'True' and the size for \textit{MaxPool1d} is $1$. $\text{Conv2d}_2$ has $12$ \textit{inchannels} and $2$ \textit{outchannels}, \textit{paddingmode} as 'replicate' and all other parameters the same as $\text{Conv2d}_1$. Note that the ReLU activation function is included after the outputs of $\text{Conv2d}_1$ and $\text{Conv2d}_2$, which are omitted in the equation above for simplicity. $\text{Conv1d}_1$ has has $1$ \textit{inchannels} and $1$ \textit{outchannels}, with \textit{kernelsize} $= 1$, \textit{padding} $= 0$, and all other parameters the same as $\text{Conv2d}_2$. 

In terms of ${\rm NN}_{\phi}^s$ in~\eqref{eq:NN_model_mu}, we used as architecture 
\[{\rm NN}_{\phi}^s =  \text{Conv1d}_2(\text{flatten}(\text{Conv2d}_4(\text{Conv2d}_3(\mathtt{input}))))\]
where the $\mathtt{input}$ contains $\big[\bs_{k-1},\, \boldsymbol{pos}^\top,\, \bq_0^\top,\, date_k \big]^\top$ ordered as a tensor in the same way as described above for the input of ${\rm NN}_{\phi}^Q$.
$\text{Conv2d}_3$ has $7$ \textit{inchannels} and $12$ \textit{outchannels}, with all other parameters the same as $\text{Conv2d}_2$ described above, while $\text{Conv2d}_4$ has all the same parameters as $\text{Conv2d}_2$. In addition, $\text{Conv1d}_2$ has the same parameters as $\text{Conv1d}_1$.
Besides, the ReLU activation function is also included in the outputs of $\text{Conv2d}_3$ and $\text{Conv2d}_4$, similar to $\text{Conv2d}_1$ and $\text{Conv2d}_2$.

For the cost function defined in \eqref{eq:loss_function_prediction_mdl}, we set the regularization parameters as $\lambda_1 = 0.1$ and $\lambda_2 = 0.001$.

\section{Supportive Experiments Results}
\label{app:SupportiveExperimentsResults}
In this section, we include additional supporting experimental results.
Figure~\ref{fig:pixelhistogram} shows the histogram representing probability of pixel values in cloudy/cloudless case among different datasets.
According to Figure~\ref{fig:pixelhistogram}, the distribution of cloudy dataset in Oroville Dam is far from that of the cloudless one, while the distribution of cloudy dataset on the Elephant Butte site is almost the same as the cloudless one. Since the outlier model adopted in this work is based on the high magnitude of outlier pixels, this means that the outlier indicator is more likely to be $1$ in Elephant Butte dataset compared to the Oroville Dam dataset, leading to a larger $\pi_k^{m,(i)}$ in~\eqref{eq:betaBernoulli}. According to~\eqref{eq:beta}, we should initialize hyperparameters as larger $e_0$ but smaller $f_0$ to achieve this goal. In summary, this shows that the hyperparameters of the robust variational image fusion framework have to be adjusted according to the difference in typical values of in-distribution and outlier pixels.

\begin{figure}
    \centering
\includegraphics[width = 1 \columnwidth]{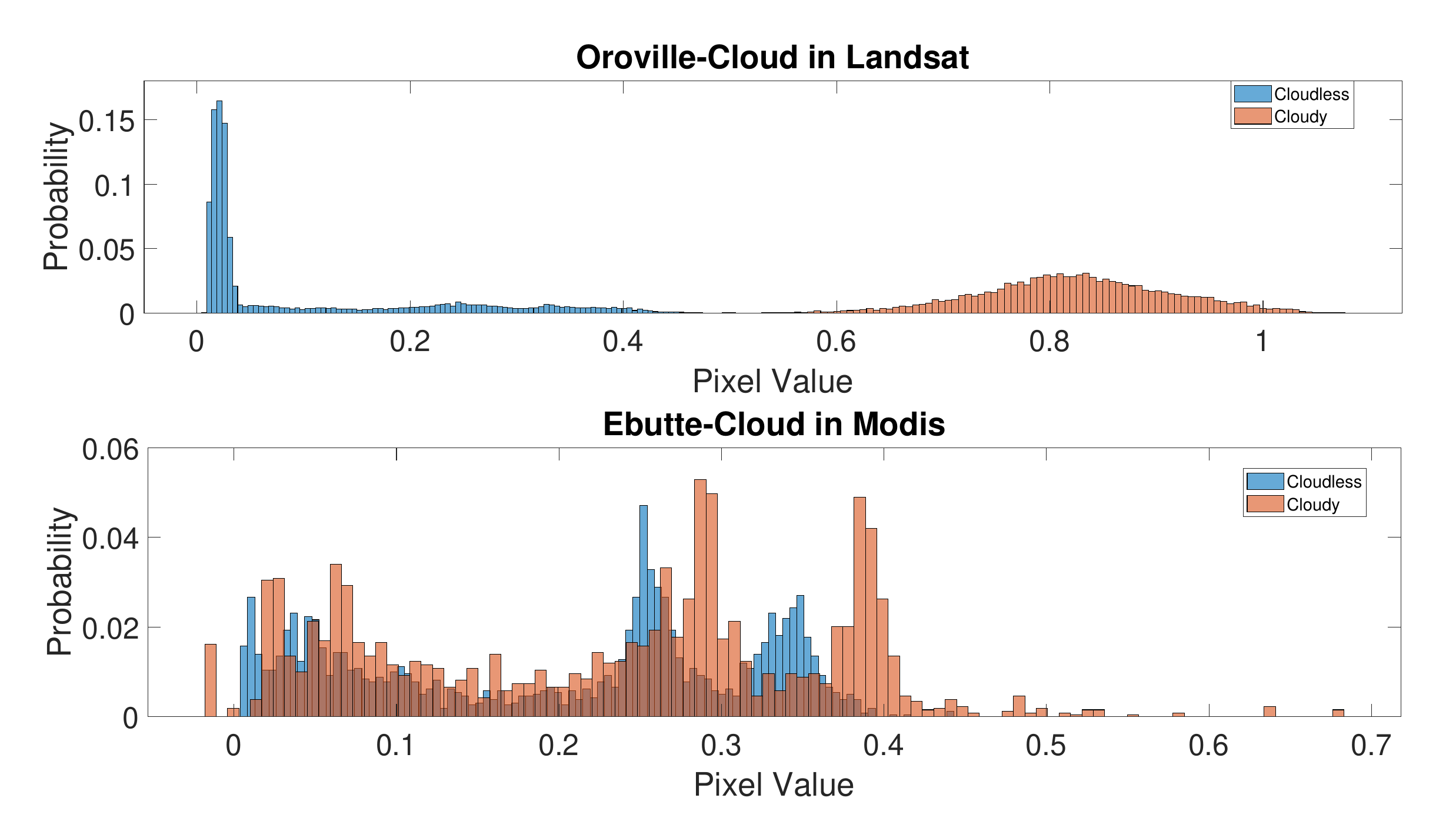}\\
\vspace{-0.3cm}
    \caption{Histogram showing the distribution of pixel values in the Oroville Dam (top panel) and Elephant Butte (lower panel) images. The blue bars represent pixel values in cloudless case while the red bars indicate pixel values in cloudy case.}
    \label{fig:pixelhistogram}
\end{figure}
Figure~\ref{fig:Reconstruction-EButte-cloudless-support} shows the fused red reflectance as well as the acquired red band reflectance values at MODIS and Landsat bands in Elephant Butte site under cloudless scenario. In the top labels, acquisition dates represent the acquired date. If the images are processed by the fusion algorithm, we have $M$ for MODIS and $L$ for Landsat, following the date in top label. As shown in the figure, only the first and last Landsat images were used in the fusion process, while the remaining two images as used as ground-truth to measure the performances of different methods. Figure~\ref{fig:watermap-Ebutte-cloudless} presents the water maps for the ground-truth (first column) and all studied algorithms obtained using K-means clustering. 

Figure~\ref{fig:Reconstruction-oroville-cloudless-support} shows the fused red and NIR band results as well as the observations and ground-truth in Oroville Dam site under cloudless scenario.
In the top labels, acquisition dates represent the acquired date. If the images are processed by the fusion algorithm, we have $M$ for MODIS and $L$ for Landsat, following the date in top label. As shown in the figure, only the first and last Landsat images were used in the fusion process, while the remaining two images as used as ground-truth to measure the performances of different methods. 
We can see that the fused images by all different methods are similar to the Landsat (ground-truth) images visually. 
The upper panel of Figure~\ref{fig:watermap-oroville-cloudless} presents the water maps for all different algorithms based on K-means clustering, with the ground truth shown at the first column, while the lower panel of Figure~\ref{fig:watermap-oroville-cloudless} shows the misclassification maps (i.e., the absolute error between the water maps by studied algorithms and the ground-truth). By comparing all the methods, we can see that their results are visually similar.

Similarly, Figure~\ref{fig:Reconstruction-oroville-cloudy-support} shows the fused red reflectance results with the acquired images and ground-truth in Oroville Dam site under cloudy scenario. 
The upper panel of Figure~\ref{fig:watermap-oroville-cloudy-interpolation} presents the water maps for all different algorithms, with the ground truth shown at the first column, based on K-means clustering, while the lower panel of Figure~\ref{fig:watermap-oroville-cloudy-interpolation}
shows the misclassification maps (i.e., the absolute error between the water maps by studied algorithms and the ground-truth). Note that the ground-truth at 05/16 in Figure~\ref{fig:watermap-oroville-cloudy-interpolation} is obtained by an interpolation method.
We can see from the results that the images estimated by KF are heavily influenced by the large cloud cover in the Landsat observation at 05/16. On the other hand, the GVIKF, KF-NN and GVIFK-NN methods hold a relatively stable performance, keeping estimated images robust against large cloud cover. 
This is also seen in the misclassification maps, where as
shown in Figure~\ref{fig:watermap-oroville-cloudy-interpolation}, the KF hold a much larger error compared with the three other methods, among which the GVIKF-NN hold the smallest misclassification amounts. 
\begin{figure*}[h!]
  \centering
  \includegraphics[width=1\linewidth]{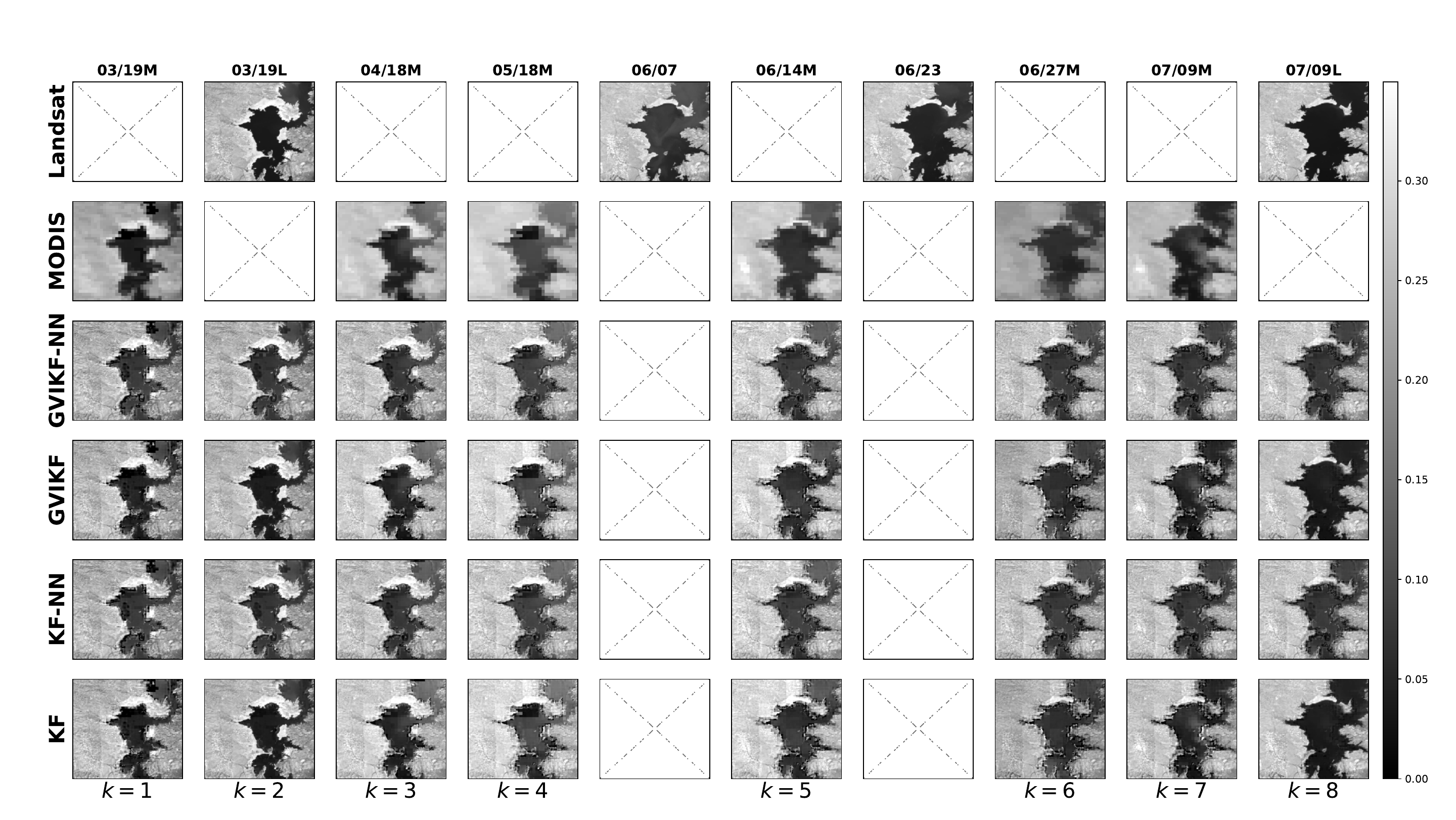}
  \label{fig:Reconstruction1-Ebutte-cloudless}
  \\\vspace{-0.6cm}
  \caption{Fusion results for the red band from MODIS (band 1) and Landsat (band 4) for the Elephant Butte example in cloudless case. The first two rows show the observed images at MODIS and Landsat bands. At each time index the fused images by GVIKF-NN, GVIKF, KF-NN and KF are presented at the following rows. Note that some Landsat images were used solely as ground-truth with only acquisition date as top labels, while images processed by the algorithms are indicated on top labels where ``M'' stands for MODIS and ``L'' for Landsat.}
  \label{fig:Reconstruction-EButte-cloudless-support}
\end{figure*}

\begin{figure*}[h]
  \centering
  \includegraphics[width=0.8\linewidth]{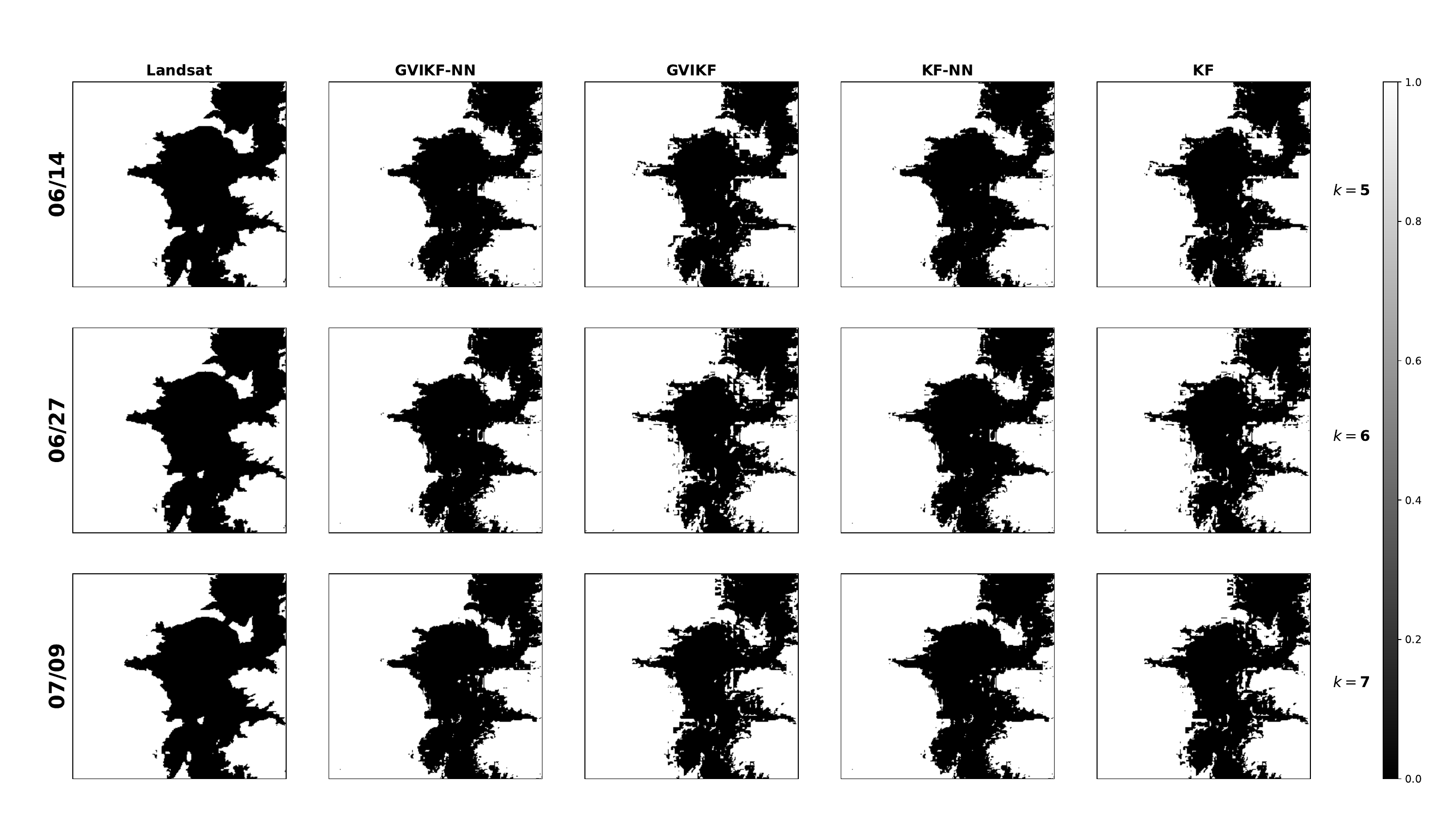}
\\\vspace{-0.4cm}
\caption{Water map of the reconstructed images in the Elephant Butte in cloudless case based on K-means clustering strategy. In the plot, 1 represents land and 0 is water pixel. Classification maps obtained from Landsat images that are not processed by the algorithms act as the ground-truth at the first column. } 
\label{fig:watermap-Ebutte-cloudless}
\end{figure*}

\begin{figure*}[h!]
  \centering
  \includegraphics[width=1\linewidth]{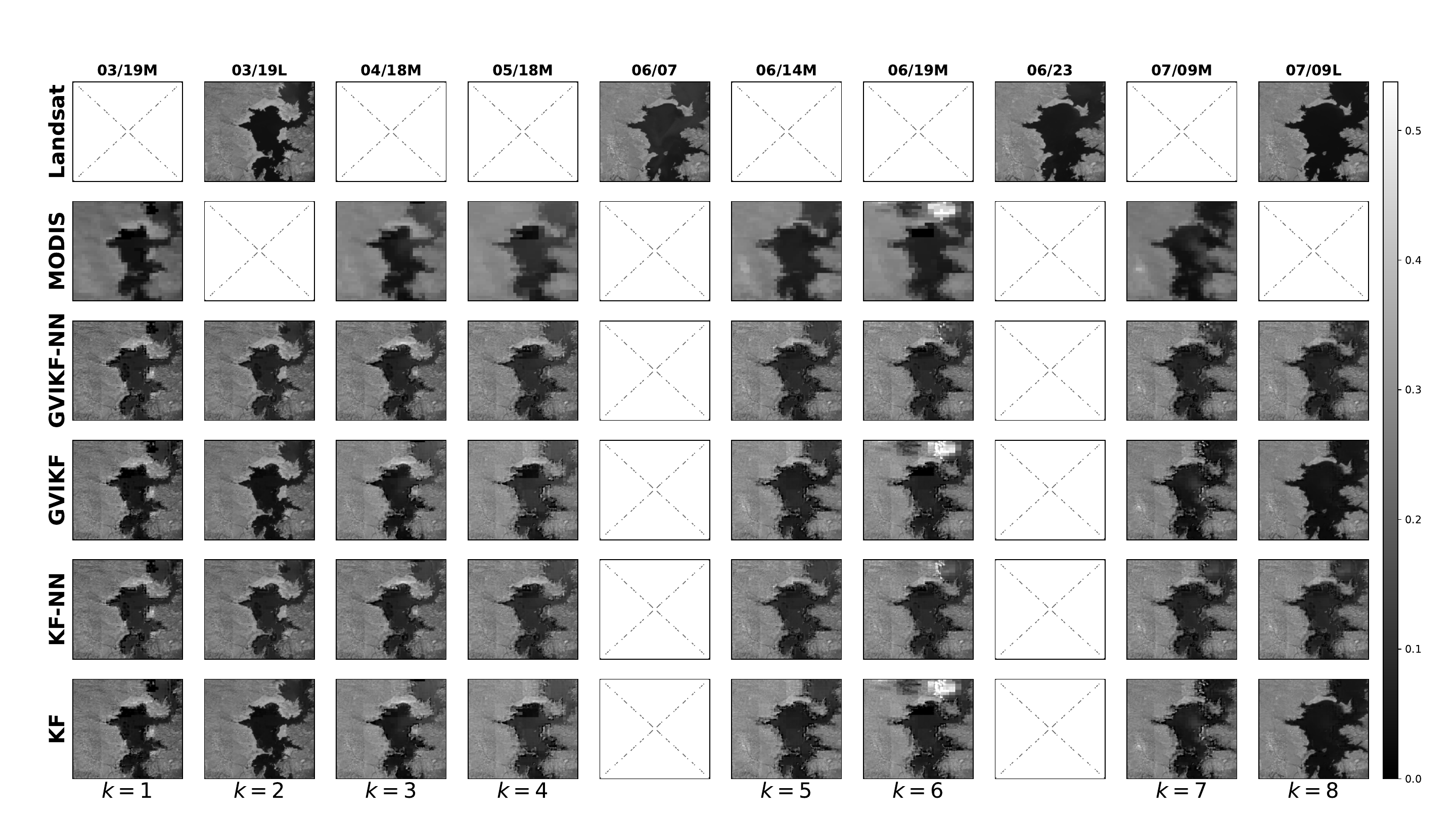}
  \label{fig:Reconstruction1-EButte-cloudy}
  \\\vspace{-0.6cm}
  \caption{Fusion results for the red band from MODIS (band 1) and Landsat (band 4) for the Elephant Butte example in cloudy case. The first two rows show the observed images at MODIS and Landsat bands. At each time index the fused images by GVIKF-NN, GVIKF, KF-NN and KF are presented at the following rows. Note that some Landsat images were used solely as ground-truth with only acquisition date as top labels, while images processed by the algorithms are indicated on top labels where ``M'' stands for MODIS and ``L'' for Landsat.}
  \label{fig:Reconstruction-EButte-cloudy-support}
\end{figure*}

\begin{figure*}[h]
  \centering
  \centering
  \includegraphics[width=0.8\linewidth]{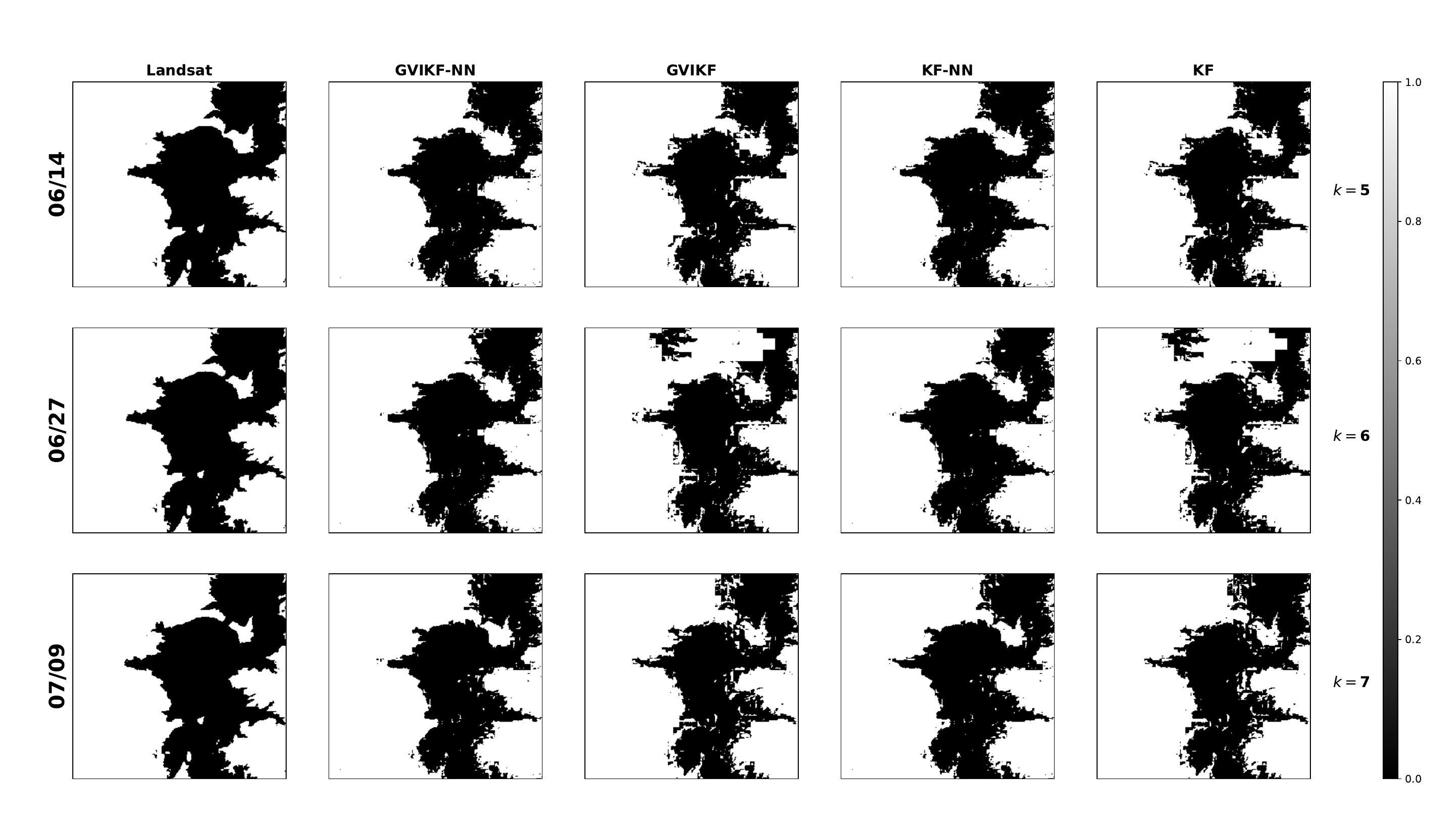}
\vspace{-0.4cm}
\caption{Water map of the fused images in the Elephant Butte in cloudy case based on K-means clustering strategy. In the plot, 1 represents land and 0 is water pixel. Classification maps obtained from Landsat images that are not processed by the algorithms act as the ground-truth at the first column. } 
\label{fig:watermap-Ebutte-cloudy}
\end{figure*}

\begin{figure*}[h!]
  \centering
  \includegraphics[width=1\linewidth]{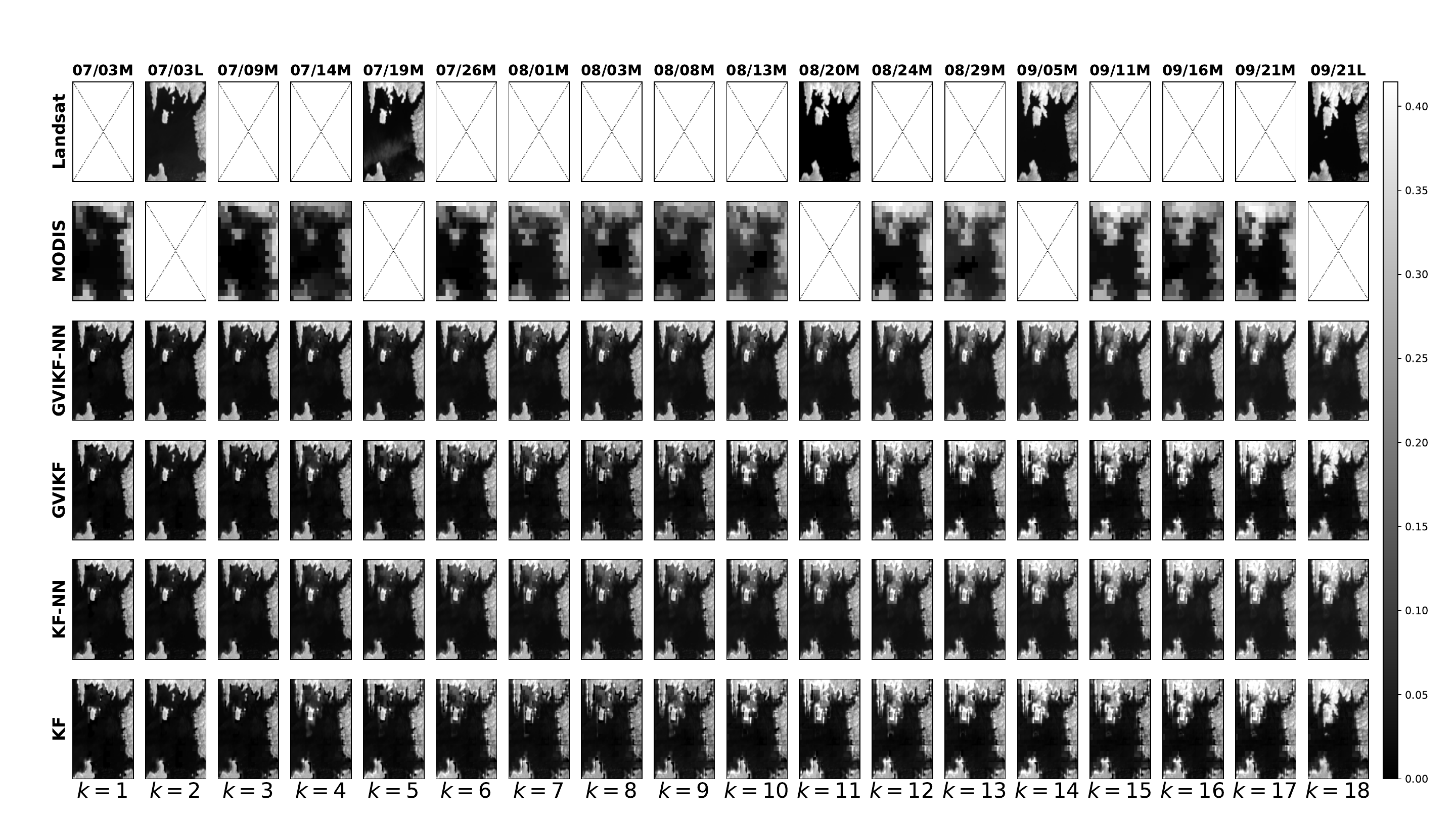}
  \includegraphics[width=1\linewidth]{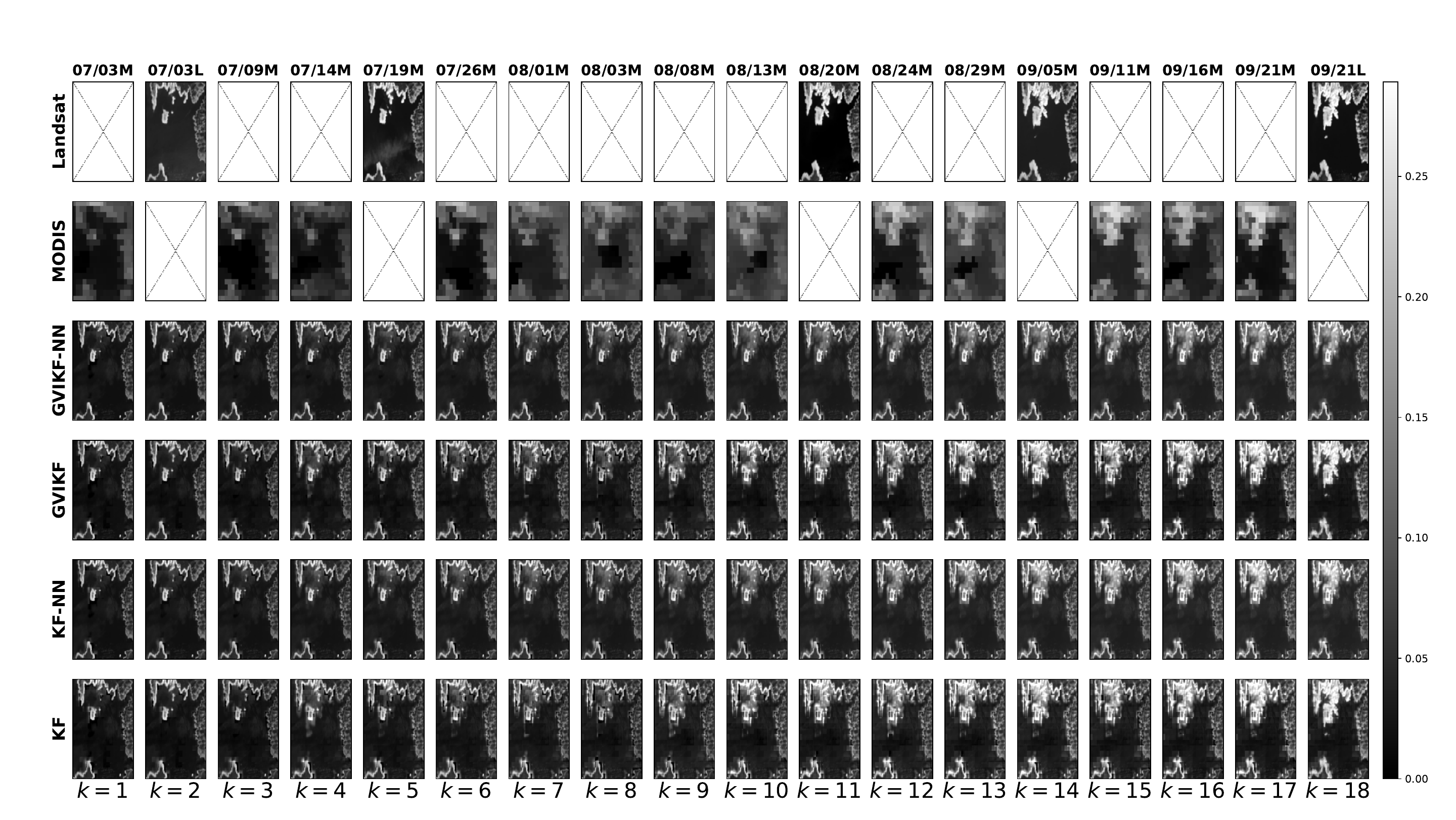}
  \caption{Fused NIR and red bands from MODIS (upper for band 2 and lower for band 1) and Landsat (upper for band 5 and lower for band 4) for the Oroville Dam example in cloudless case using different strategies. The first two rows of the top and bottom subfigures show the observed images at MODIS and Landsat bands. At each time index the fused images given by GVIKF-NN, GVIKF, KF-NN and KF are presented at the following rows. Note that some Landsat images were used solely as ground-truth with only acquisition date as top labels, while images used are indicated on top labels where ``M'' stands for MODIS and ``L'' for Landsat.}\label{fig:Reconstruction-oroville-cloudless-support}
\end{figure*}

\begin{figure*}[h]
  \centering
  \includegraphics[width=0.8\linewidth]{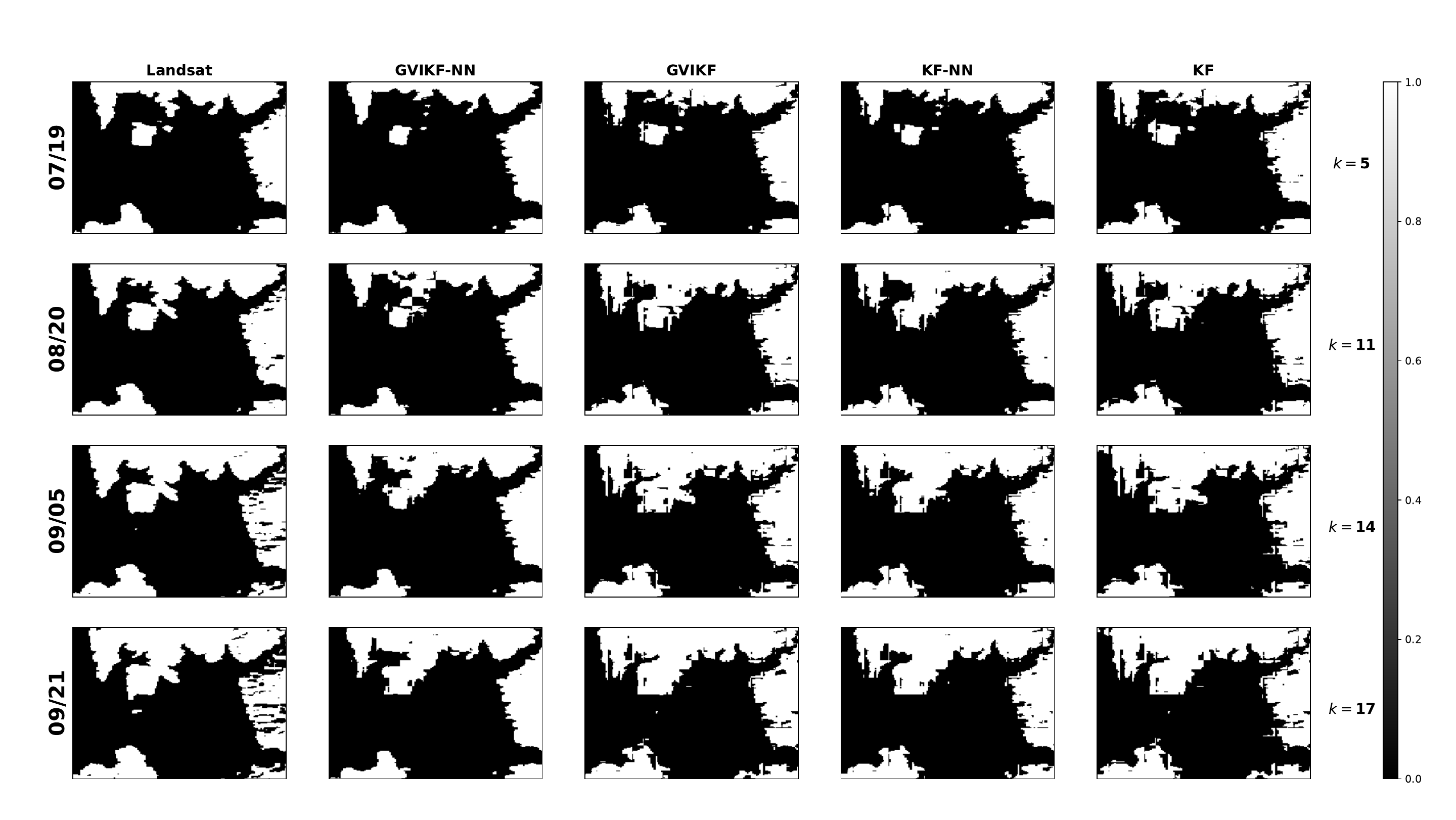}
  \centering
  \includegraphics[width=0.8\linewidth]{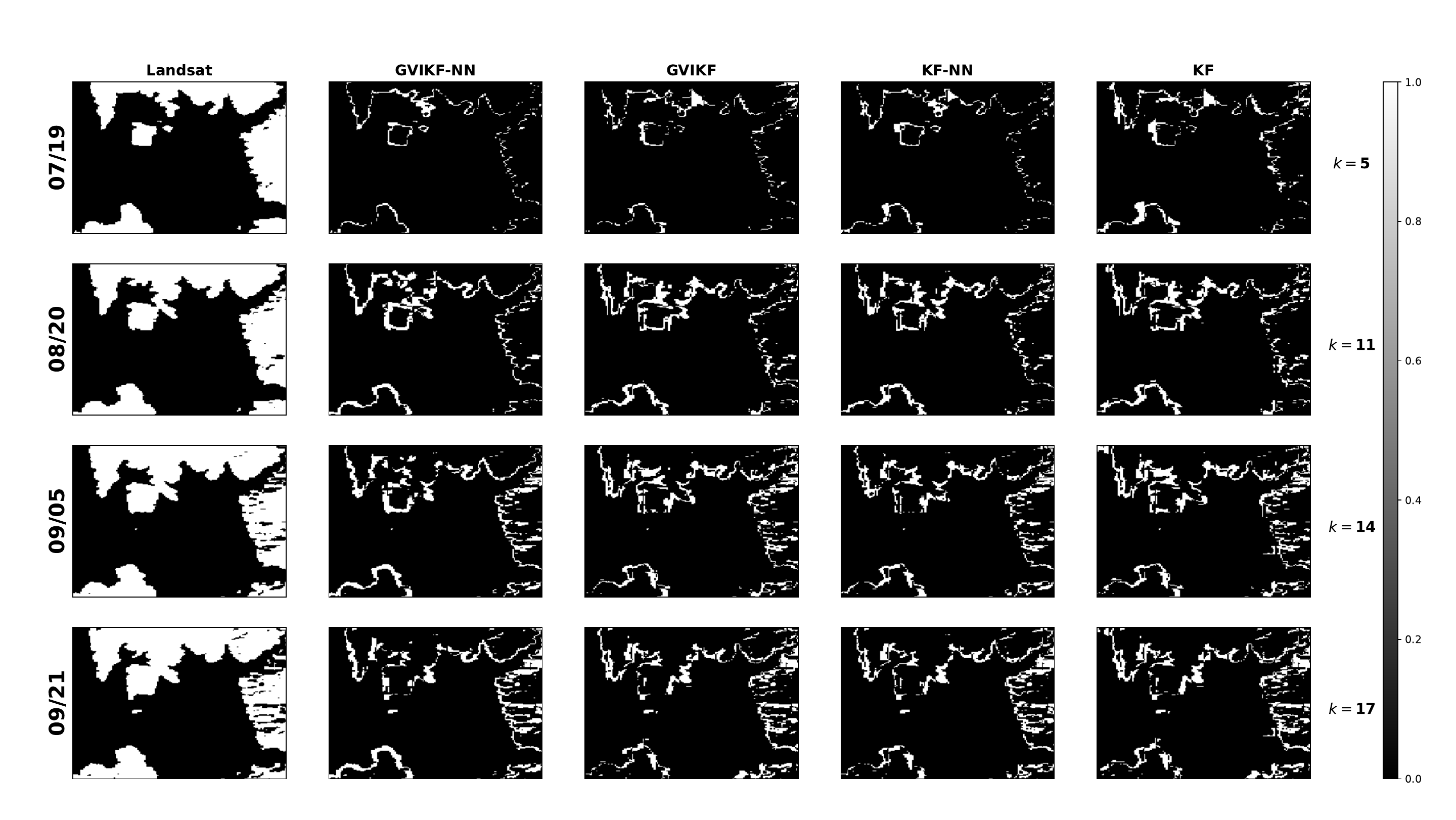}
\vspace{-0.4cm}
\caption{(\textbf{Upper Panel}) Water map of the fused results in the Oroville Dam in cloudless case based on K-means clustering strategy. In the plot, 1 represents land and 0 is water pixel. Classification maps obtained from Landsat images that are not processed by the algorithms act as the ground-truth at the first column. (\textbf{Lower Panel}) Absolute error of water map of images based on K-means clustering strategy. In this plot, 0 pixel value indicates correct classification and 1 pixel value indicates misclassification. For comparison, acquired Landsat images are shown in the first column as the ground-truth.} 
\label{fig:watermap-oroville-cloudless}
\end{figure*}
\begin{figure*}[h!]
  \centering
  \includegraphics[width=1\linewidth]{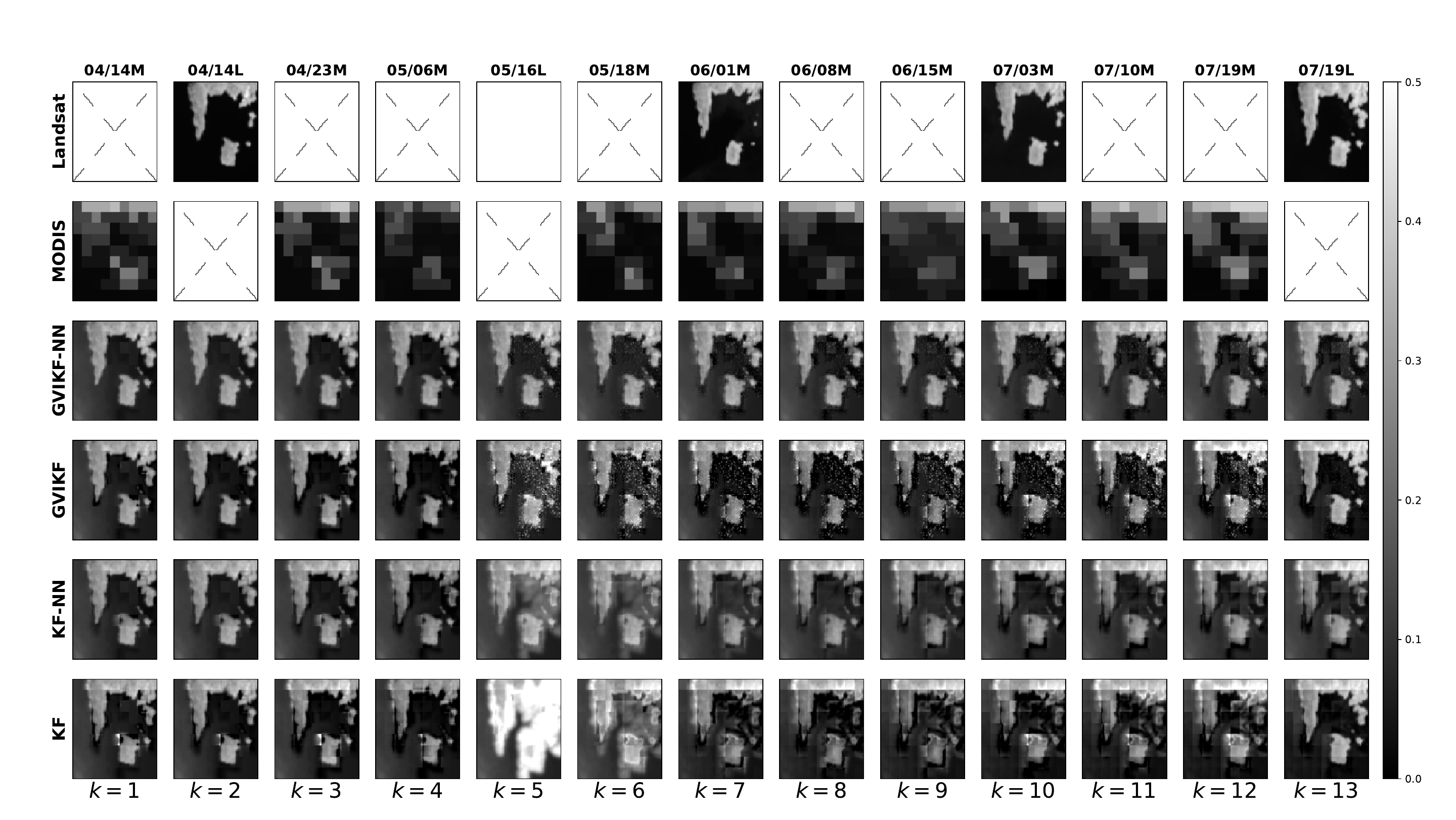}
  \includegraphics[width=1\linewidth]{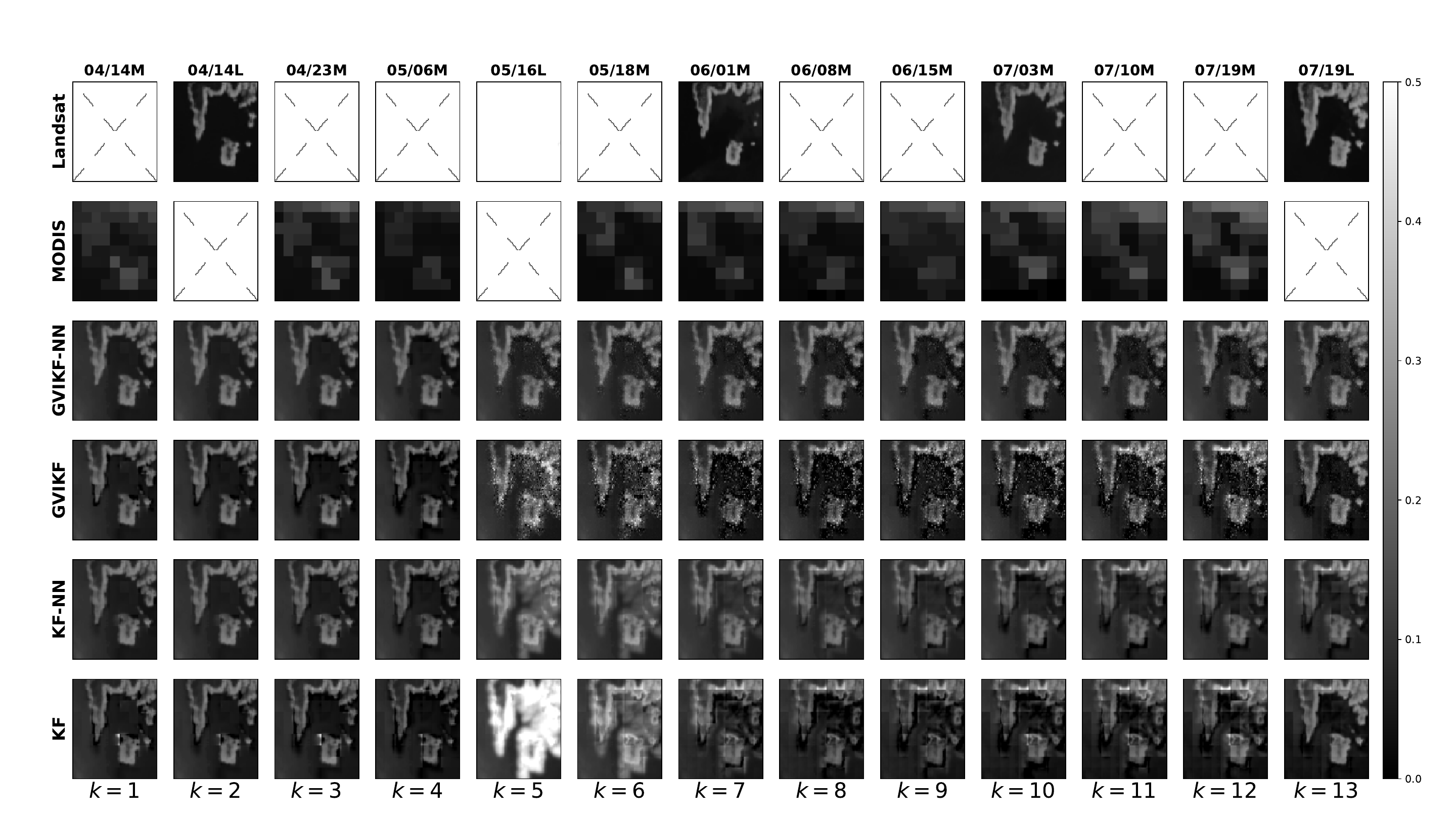}
  \caption{Fused NIR and red bands from MODIS (upper for band 2 and lower for band 1) and Landsat (upper for band 5 and lower for band 4) for the Oroville Dam example in the cloudy case using different strategies. The first two rows of the top and bottom subfigures show the observed images at MODIS and Landsat bands. At each time index the fused images by GVIKF-NN, GVIKF, KF-NN and KF are presented at the following rows. Note that some Landsat images were used solely as ground-truth with only acquisition date as top labels, while images used are indicated on top labels where ``M'' stands for MODIS and ``L'' for Landsat.}\label{fig:Reconstruction-oroville-cloudy-support}
\end{figure*}
\begin{figure*}[h]
  \centering
  \includegraphics[width=0.8\linewidth]{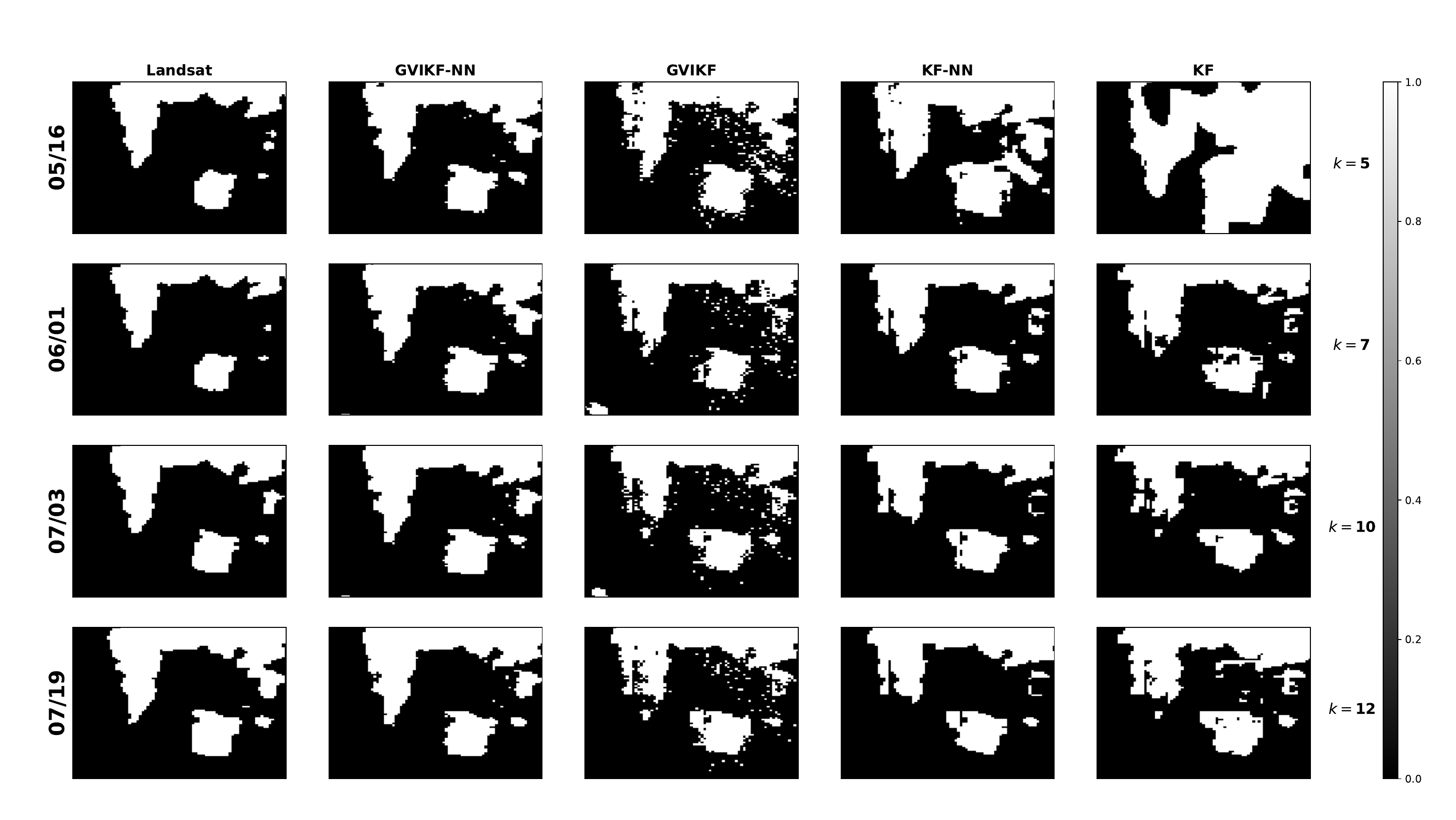}
  \centering
  \includegraphics[width=0.8\linewidth]{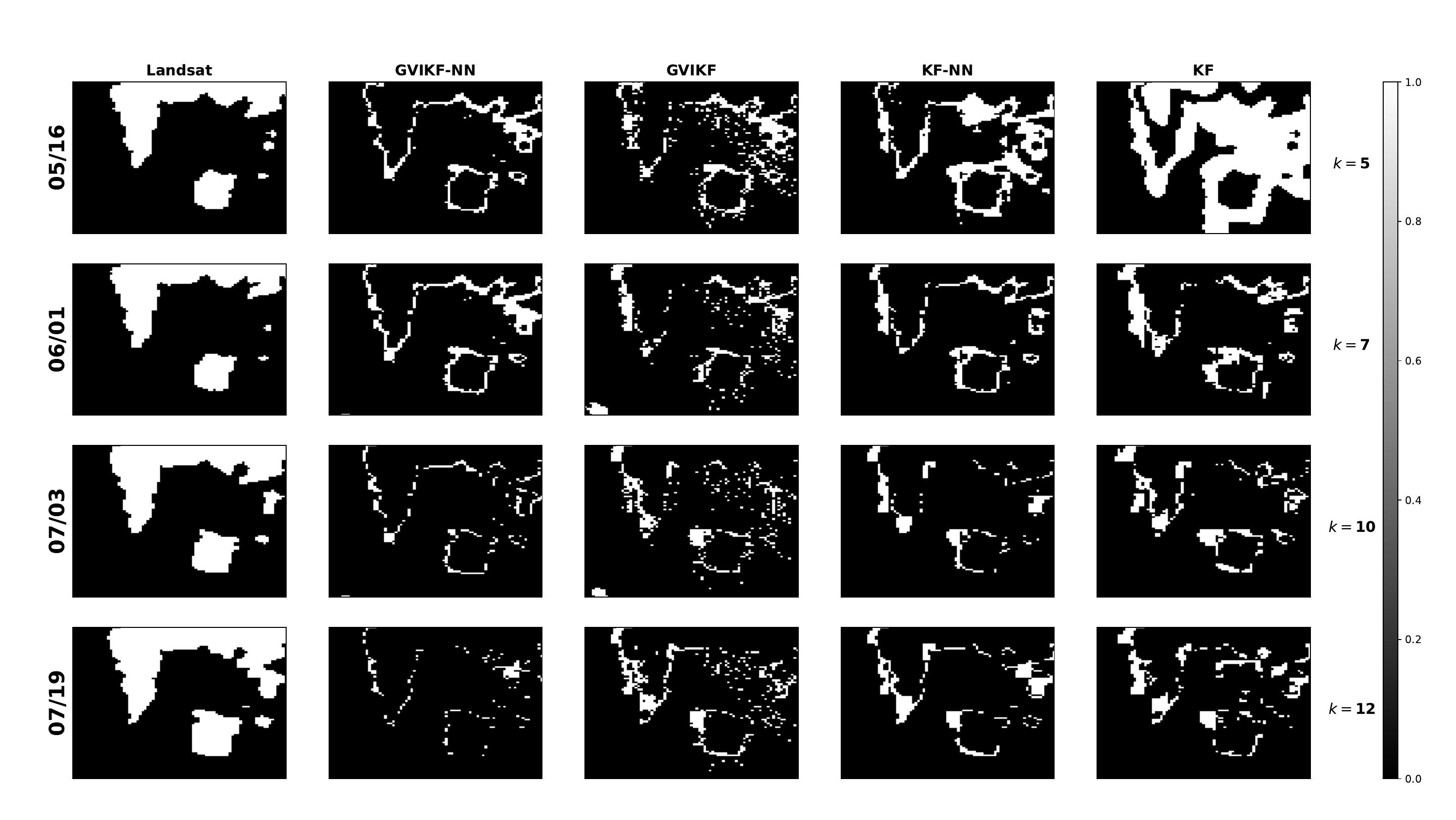}
\vspace{-0.4cm}
\caption{(\textbf{Upper Panel}) Water map of the fused images in the Oroville Dam example in cloudy case based on K-means clustering strategy. In the plot, 1 represents land and 0 is water pixel. Classification maps obtained from Landsat images that are not processed by the algorithms act as the ground-truth at the first column. (\textbf{Lower Panel}) Absolute error of water map of images based on K-means clustering strategy. In this plot, 0 pixel value indicates correct classification and 1 pixel value indicates misclassification. For comparison, acquired Landsat images are shown in the first column as the ground-truth (the Landsat ground truth at 05/16 is a surrogate one obtained using an interpolation strategy).} 
\label{fig:watermap-oroville-cloudy-interpolation}
\end{figure*}

\clearpage
\bibliographystyle{elsarticle-num}
\bibliography{bibo}
\end{document}